\documentclass[]{aastex631}

\begin{document}

\title{Commissioning of the MIRAC-5 Mid-Infrared Instrument on the MMT}

% \author{Anonymous}

\author[0000-0003-0949-7212]{Rory Bowens}
\affiliation{Department of Astronomy, University of Michigan, 1085 S University Avenue, Ann Arbor, MI, 48103 USA}

\author[0000-0002-0834-6140]{Jarron Leisenring}
\affiliation{Steward Observatory and Department of Astronomy, University of Arizona, 933 N Cherry Avenue, Tucson, AZ, 85721 USA}

\author[0000-0003-1227-3084]{Michael R. Meyer}
\affiliation{Department of Astronomy, University of Michigan, 1085 S University Avenue, Ann Arbor, MI, 48103 USA}

\author[0000-0001-8103-5499]{Taylor L. Tobin}
\affiliation{Department of Astronomy, University of Michigan, 1085 S University Avenue, Ann Arbor, MI, 48103 USA}

\author{Alyssa L. Miller}
\affiliation{Department of Astronomy, University of Michigan, 1085 S University Avenue, Ann Arbor, MI, 48103 USA}

\author{John D. Monnier}
\affiliation{Department of Astronomy, University of Michigan, 1085 S University Avenue, Ann Arbor, MI, 48103 USA}

\author{Eric Viges}
\affiliation{Space Physics Research Laboratory, University of Michigan, 1085 S University Avenue, Ann Arbor, MI, 48103 USA}

\author{Bill Hoffmann}
\affiliation{Steward Observatory and Department of Astronomy, University of Arizona, 933 N Cherry Avenue, Tucson, AZ, 85721 USA}

\author{Manny Montoya}
\affiliation{Steward Observatory and Department of Astronomy, University of Arizona, 933 N Cherry Avenue, Tucson, AZ, 85721 USA}

\author{Olivier Durney}
\affiliation{Steward Observatory and Department of Astronomy, University of Arizona, 933 N Cherry Avenue, Tucson, AZ, 85721 USA}

\author{Grant West}
\affiliation{Steward Observatory and Department of Astronomy, University of Arizona, 933 N Cherry Avenue, Tucson, AZ, 85721 USA}

\author{Katie Morzinski}
\affiliation{Steward Observatory and Department of Astronomy, University of Arizona, 933 N Cherry Avenue, Tucson, AZ, 85721 USA}

\author{William Forrest}
\affiliation{Department of Physics and Astronomy, University of Rochester, 206 Baucsh and Lomb Hall, Rochester, NY, 14627 USA}

\author{Craig McMurtry}
\affiliation{Department of Physics and Astronomy, University of Rochester, 206 Baucsh and Lomb Hall, Rochester, NY, 14627 USA}



\begin{abstract}

%300 word limit; the following text is from the SPIE MIRAC-5 proposal
%intro
We present results from commissioning observations of the mid-IR instrument, MIRAC-5, on the 6.5-m MMT telescope. MIRAC-5 is a novel ground-based instrument that utilizes a state-of-the-art GeoSnap (2 - 13 microns) HgCdTe detector with adaptive optics support from MAPS to study protoplanetary disks, wide-orbit brown dwarfs, planetary companions in the contrast-limit, and a wide range of other astrophysical objects.
%work
We have used MIRAC-5 on six engineering observing runs, improving its performance and defining operating procedures.
%results
We characterize key aspects of MIRAC-5's performance, including verification that the total telescope, atmosphere, instrument, and detector throughput is approximately 10\%. Following a planned dichroic upgrade, the system will have a throughput of 20\% and background limiting magnitudes (for SNR = 5 and 8 hour exposure times) of 18.0, 15.6, and 12.6 for the L', M', and N' filters, respectively. The detector pixels experience 1/f noise but, if the astrophysical scene is properly modulated via chopping and nodding sequences, it is less than 10\% the Poisson noise from the observed background in an 85 Hz frame. We achieve close to diffraction-limited performance in the N-band and all bands are expected to reach diffraction-limited performance following the adaptive optics system commissioning. We also present an exposure time calculator calibrated to the on-sky results.
%impact
In its current state, MIRAC-5 will be capable of achieving several scientific objectives including the observation of warm wide-orbit companions. Once the adaptive optics is commissioned and a coronagraph installed in 2025, MIRAC-5 will have contrast-limited performance comparable to JWST, opening new and complementary science investigations for close-in companions.

\end{abstract}

\keywords{Exoplanets (X) --- Mid-infrared astronomy(X)}

\section{Introduction} 
\label{sec:intro}
    % Importance of direct imaging (scientific values for different objects)
   Direct imaging of exoplanets and brown dwarf companions can be used to determine fluxes at multiple wavelengths (which, combined with measured luminosities and effective temperature, leads to radii), constrain atmospheric compositions (based on emission spectra), and test planetary evolution theory \citep{traub_oppen_2010}.
    % Focus on mid-IR direct imaging
    Mid-infrared (mid-IR, 3 - 13 microns) direct imaging is particularly relevant to warm (300 - 1000 K) systems which emit the bulk of their energy in those wavelengths. Mid-IR light penetrates obscuring dust and we can use it to probe warm rocky and gaseous worlds with less stringent contrast requirements compared to reflected light observations at shorter wavelengths \citep{pathak2021arXiv210413032P}.

    %mid-IR ground-based development has been modest since the launch of Spitzer
    In 2003, the launch of the Spitzer Space  Telescope \citep{spitzer2004ApJS..154....1W} led to significant advances in mid-IR astronomy. Unlike ground-based systems, space-based observatories are not impacted by atmospheric thermal emission and have the capacity to be cooled to substantially reduce background emission from otherwise warm telescope optics and structures, eliminating one of the most significant sources of noise of mid-IR observations. Furthermore, space-based observatories do not suffer from atmospheric absorption which precludes observations of certain bandpasses from the ground. The success of Spitzer and other infrared space missions such as ISO, WISE, and the ongoing JWST has led to reduced efforts on ground-based mid-IR instrumentation. But emerging technology and facility improvements are reopening the discovery space from the ground. New infrared detectors, such as Teledyne Imaging Sensor's (TIS) GeoSnap, offer deep wells, fast readouts, and, most importantly, high quantum efficiency (QE) and low noise within the 2 to 13 micron (half peak QE cutoffs) regime \citep{jarron2023AN....34430103L} needed for ground-based observations to overcome the high background rates from thermal emission of the sky and telescope. Additionally, new advanced adaptive optics (AO) systems on $>$ 6.5 meter telescopes enable superior angular resolution to current space-based systems at all wavelengths \citep{leisenring2012SPIE.8446E..4FL, davies2012ARA&A..50..305D}. In the context of exoplanet studies, ground-based instruments coupled with AO systems with Strehl $>$ 90\% \citep{leisenring2012SPIE.8446E..4FL, davies2012ARA&A..50..305D} can also achieve greater contrast-limited performance than space-based instruments \citep[see work done by the Near Earths in the Alpha Cen Region (NEAR) project, e.g;][]{wagner_2021, beichmann_2020, guyon_2018, pathak2021arXiv210413032P}, opening a window into habitable zone terrestrial planets \citep{bowens2021A&A...653A...8B}. These factors motivate plans for future mid-IR instruments in era of extremely large telescopes (ELT) such as the Mid-infrared ELT Imager and Spectrograph (METIS) on the European ELT \citep{brandl202410.1117/12.3018975}.

    % - history of MIRAC (and BLINC). 
% - and now we have Geosnap too. 
    Ahead of the deployment of these detectors on ELTs, the fifth Mid-Infrared Array Camera (MIRAC-5) instrument on the MMT 6.5-meter telescope will be one of the only AO supported ground-based mid-infrared imagers in the world. MIRAC was first developed in 1988 as a collaboration between the University of Arizona (UA), the Smithsonian Astrophysical Observatory, and the Naval Research Laboratory \citep{HOFFMANN1994175, hoffmann1998SPIE.3354..647H}. The system has gone through several iterations over the past three decades with evolving science objectives. MIRAC-3, which operated from 2000 to 2008 on MMT and Magellan, was mated with the Bracewell Infrared Nulling Cryostat (BLINC) as a prototype to LBTI's Nulling and Imaging Camera (NIC) \citep{hinz2000SPIE.4006..349H, hinz10.1117/12.790242}. While the nulling interferometry is no longer used, BLINC is still mated to the MIRAC system for reimaging purposes. The latest iteration, MIRAC-5, is a collaboration between the University of Michigan (UM) and UA. It features a 13-$\mu$m cut-off HgCdTe (MCT) GeoSnap detector supplied by TIS and the University of Rochester \citep{bowens2022SPIE12184E..1UB, jarron2023AN....34430103L}. The GeoSnap has deep well depths (1.2 million e-) and fast readouts. Compared to prior mid-IR ground-based detectors, GeoSnap offers wider wavelength coverage, reduced 1/f noise, and a larger format (with the MIRAC-5 GeoSnap being 1024x1024 and future GeoSnap being 2048x2048 format). With AO support from the MMT Adaptive optics exoPlanet characterization System (MAPS) \citep{morzinski2020SPIE11448E..1LM}, MIRAC-5 will be capable of imaging wide orbit companions and potentially detecting key molecules in their atmospheres such as NH$_3$ \citep{bowens2022SPIE12184E..1UB}. A planned annular groove phase mask coronagraph \citep[AGPM;][]{mawet2005ApJ...633.1191M} upgrade should enable 2 to 10 $\lambda/D$ (contrast-limited regime) capabilities similar to NEAR \citep{wagner_2021} which achieved 3$\sigma$ N-band contrasts of $3\times10^{-6}$ for Alpha Cen A within 1 to 1.5 arcseconds. This contrast-limited performance is comparable to MIRI's on JWST which achieved 3$\sigma$ contrasts of $2-4\times10^{-5}$ within 1 arcsecond for the F1065C and F1140C filters using principal component analysis subtraction of reference stars (\cite{boccaletti2022A&A...667A.165B}, c.f., \cite{godoy2024arXiv240903485G}).
    
    %State of instrument
    MIRAC-5 has been used during five observing runs in concert with MAPS engineering and commissioning. Here, we summarize key capabilities of the system as well as expected improvements to image quality once coupled with the fully commissioned MAPS AO. We focus particularly on a comparison of the on-sky performance of MIRAC-5 with its expected performance and comparable systems.
    %Other ground-based infrared systems exhibit differences between predicted performance estimates and observations \citep[e.g.,][]{bowensrubin2023AJ....166..260B} which may be due to non-Poissonian sky noise (such as water vapor variations). We plan to verify if MIRAC-5 is impacted by these non-Poissonian noise sources.

    % Give paper outline
    We begin with a description of the MIRAC-5 instrument and the overall optical design in Section \ref{sec:instrument}. We then discuss performance verification of the instrument in Section \ref{sec:performance} followed by an explanation and demonstration of an exposure time calculator for MIRAC-5 in Section \ref{sec:program}. We compare performance relative to other mid-IR instruments and discuss future plans in Section \ref{sec:discussion}, before concluding in Section \ref{sec:conclusion}.

\section{Overview of the Instrument} 
\label{sec:instrument}
    % Keep it short and to the point
    We define the system in three parts: the MIRAC-5 Instrument, the GeoSnap detector, and the host MMT telescope supported by MAPS AO. We conclude this section by explaining the instrument control and operational modes.

    \subsection{MIRAC-5 Instrument}
\begin{figure}
        \centering
        \includegraphics[width=0.5\linewidth]{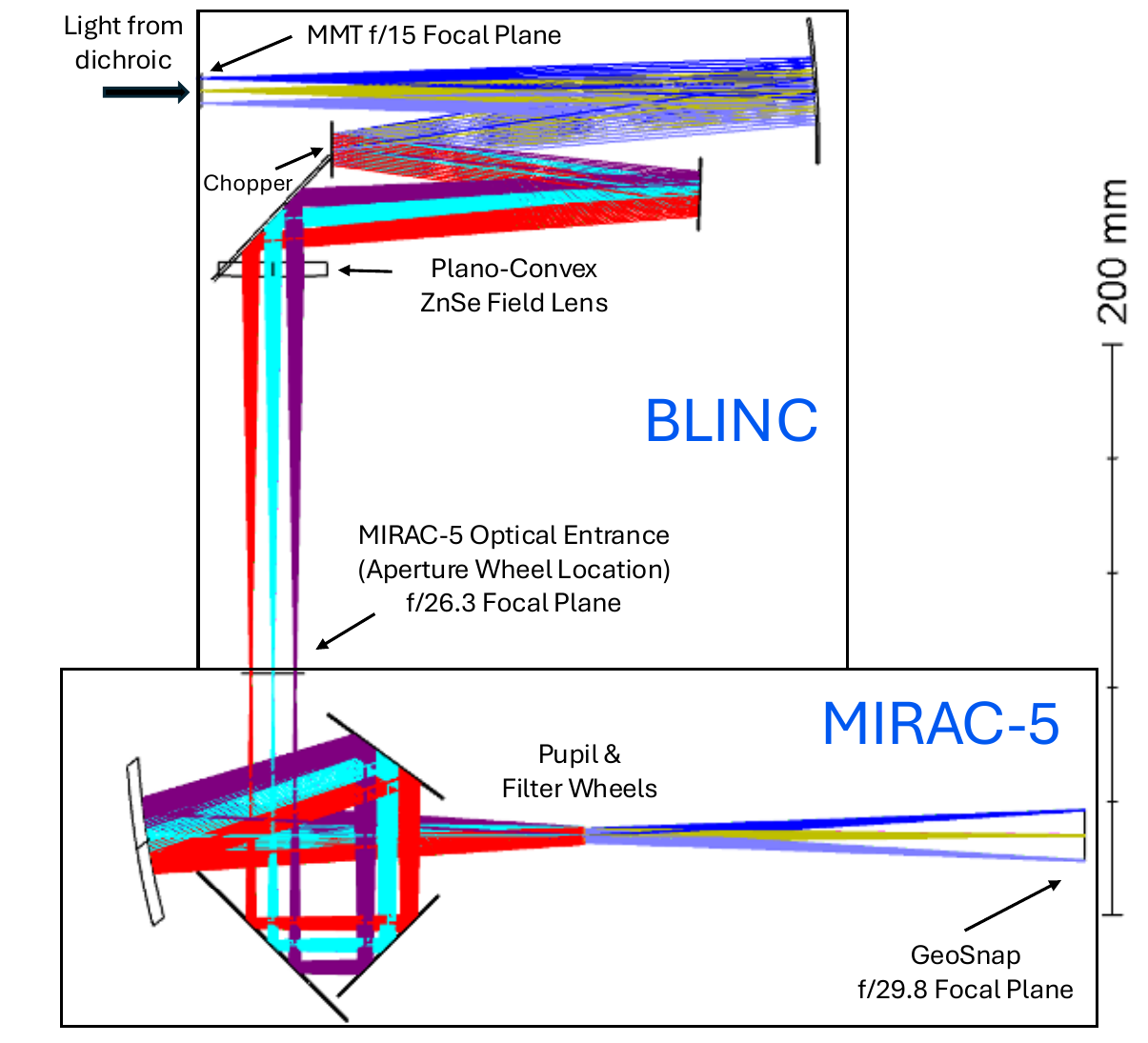}
        \caption{An updated version of BLINC and MIRAC-5 optical layout originally presented in \citet{bowens2022SPIE12184E..1UB}. Minor adjustments to the focal plane have altered the intermediate and final focal plane f/\#s.}
        \label{fig:optical_layout}
    \end{figure}

    The layout of the MIRAC-5 cryostat is covered in \citet{bowens2022SPIE12184E..1UB}. MIRAC-5 is pulse-tube cooled and maintains the enclosed GeoSnap detector at a user selected operating temperature between 35 and 45 K with a precision of $<$ 20 mK (via close-loop heater control). MIRAC-5 interfaces to the BLINC cryostat, a liquid nitrogen cooled system at 77 K used for reimaging the Cassegrain focus of the MMT adaptive secondary through MIRAC-5 optics and then onto the detector. Telescope light enters BLINC via a KRS-5 dewar window and then passes through a windowless interface to the MIRAC-5 cryostat. We present a modified version of the optical layout given in Figure 3 from \citet{bowens2022SPIE12184E..1UB} in Figure \ref{fig:optical_layout}. Together, the two cryostats reimage the f/15 beam from the MMT adaptive secondary telescope to an f/29.8 focal plane at the detector.

    In the optical path is an articulating flat mirror (e.g., pupil plane chopping mirror) driven by a rotational voice coil actuator capable of moving the source by approximately 7.4 arcseconds (410 pixels) along the horizontal axis. An aperture wheel is located at an intermediate focal plane near the the MIRAC-5 optical entrance with a temperature of 18 K. A cold pupil wheel and two filter wheels are located between this intermediate focal plane and the GeoSnap. Although these wheels are covered in \citet{bowens2022SPIE12184E..1UB}, their positions are updated here in Tables \ref{tab:pasp_apwheel}, \ref{tab:pasp_pupwheel}, and \ref{tab:pasp_filtwheel} as several filters have been changed.

    \begin{table}[ht]
        \centering
        \begin{tabular}{c|c|c}
        Number     &  Name & Size\\
        \hline
        1 (Home)     & Large Square & 25.4x25.4 mm \\
        2 & Pinhole & 0.5 mm \\
        3 & Large Slit - Center & 0.71x25.4 mm \\
        4 & Small Slit - Offset & 0.58x25.4 mm \\
        5 & Large Slit - Offset & 0.71x25.4 mm \\
        \hline
        \end{tabular}
        \caption{Aperture wheel positions. The aperture wheel is located at the entrance of MIRAC-5.}
        \label{tab:pasp_apwheel}
    \end{table}
    
    \begin{table}[ht]
        \centering
        \begin{tabular}{c|c|c}
        Number     & Name & Size \\
        \hline
        1 (Home)     & Pinhole & 0.51 mm \\
        2 & Magellan & 8.99 mm \\
        3 & Dual Pupil - Nominal & 2 x 1.78 mm \\
        4 & MMT & 6.08 mm \\
        5 & Dual Pupil - Oversize & 2 x 1.96 mm \\
        6 & Blank & \\
        \hline
        \end{tabular}
        \caption{Pupil wheel settings. The pupil wheel and the two filter wheels are located in series before the detector plane.}
        \label{tab:pasp_pupwheel}
    \end{table}
    
    \begin{table}[ht]
        \centering
        \begin{tabular}{c|c|c|c|c|c|c}
         Number    & Name & Central $\lambda$ ($\mu m$) & FWHM ($\mu m$)& T$_{filter}$ & Current T$_{dichroic}$ & Future T$_{dichroic}$\\
         \hline
         W1-1 (Home)   & Open & & &&& \\
         W1-2	&L’	&3.84	&0.62	&0.90 &0.57 & 0.95\\
        W1-3	&N0790	&7.93	&0.70	&0.87 &0.48 & 0.99\\
        W1-4	&W0870	&8.74	&1.23	&0.89 & 0.49 & 0.99\\
        W1-5	&N0915	&9.15	&0.80	&0.86 & 0.50 & 0.99\\
        W1-6	&N0980	&9.82	&0.92	&0.86 & 0.51 & 0.98\\
        W1-7	&W1055	&10.57	&0.97	&0.90 & 0.51 & 0.99\\
        W1-8	&Ammonia	&10.59	&0.64	&0.87 & 0.51 & 1.00\\
        W1-9	&N1185	&11.89	&1.14	&0.84 & 0.43 & 0.95\\
        W1-10	&N1252	&12.55	&1.17	&0.83 & 0.43 & 0.97\\
        W1-11	&N’	&11.34	&2.27	&0.87 & 0.47 & 0.97\\
        W1-12  &Blank &&&&& \\
        \hline
        W2-1 (Home)   & Open & & &&& \\
        W2-2	&N-band	&10.85	&5.71	&0.89 & 0.48 & 0.98 \\
        W2-3	&BaF2 Blocker*&&&&&\\
        W2-4	&EO 64355**	&&&&&\\
        W2-5	&H-band	&1.65	&0.33	&0.99 & 0.34 & 0.004\\
        W2-6	&K-band	&2.22	&0.35	&0.61 & 0.40 & 0.92\\
        W2-7   &Blank &&&&& \\
        W2-8	&M’	&4.66	&0.24	&0.78 & 0.32 & 0.92\\
        W2-9	&M-band	&4.76	&0.59	&0.90 & 0.29 & 0.92\\
        W2-10	&PIL (New)		&&&	&&\\		
        W2-11	&PIL (Old)		&&&	&&\\		
        W2-12	&Blank	&&&&&\\
        
        \hline
        \end{tabular}
        \caption{Filter Wheels 1 and 2 labeled with the central wavelengths (Central $\lambda$), the full-width half-maximum (FWHM), the filter transmission (median transmission within the FWHM), the median dichroic transmission (current dichroic) within the FWHM, and the median dichroic transmission (for a newly purchased dichroic) within the FWHM. The Ammonia (Amm.) filter is a narrow bandpass filter optimized for the detection of the 10.6 micron ammonia feature \citep{bowens2022SPIE12184E..1UB}. The Pupil Imaging Lenses (PIL) can be used to focus the pupil wheel on the detector for alignment purposes and comes in 25.4 mm diameter (new) or 19.0 mm diameter (old). *Blocking at greater than 10 microns. **Neutral density filter with 10\% throughput in observable wavelengths.}
        \label{tab:pasp_filtwheel}
    \end{table}

    \subsection{GeoSnap}
    \label{sec:geosnap_desc}
\begin{figure}
        \centering
        \includegraphics[width=0.5\linewidth]{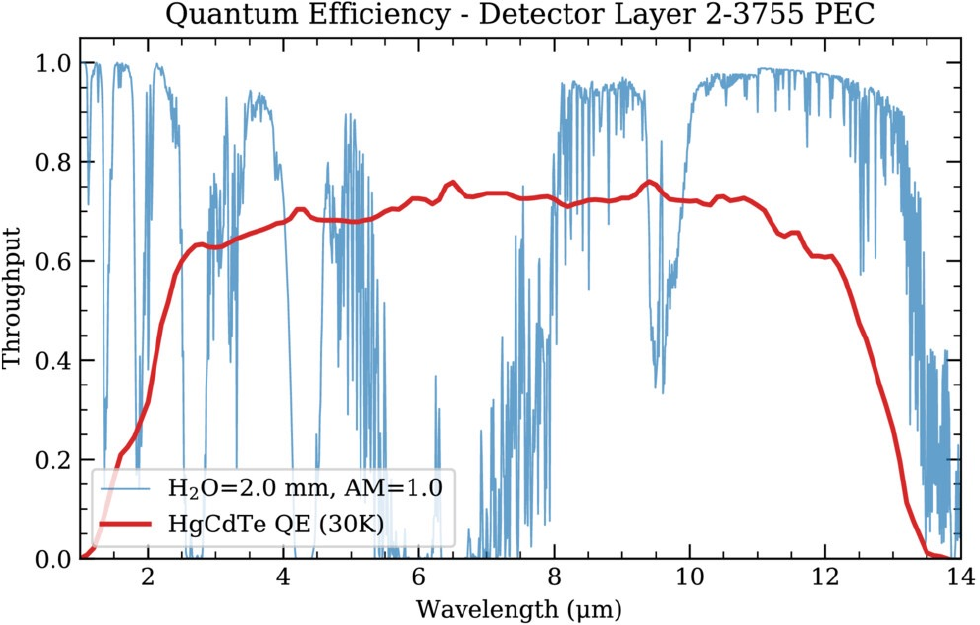}
        \caption{The quantum efficiency of the MIRAC-5 GeoSnap as determined by TIS via a sample of the photosensitive material \citep{jarron2023AN....34430103L}. Modelled atmospheric transmission is provided for comparison \citep{lord1992}.}
        \label{fig:geoqe}
    \end{figure}

    The MIRAC-5 GeoSnap was characterized in \citet{jarron2023AN....34430103L} and its major properties are summarized in Table \ref{tab:properties}. The MIRAC-5 GeoSnap is an MCT detector from TIS which is sensitive from 2 to 13 microns (half peak QE cutoffs) with an average QE of 65\% (without an anti-reflection coating) and an active quadrant of 1024x1024 pixels (18-micron pixel pitch). It has limited quantum efficiency shortward to 1 micron as seen in Figure \ref{fig:geoqe}. The detector has a well depth of 1.2 million electrons and a readout integrated circuit (ROIC) capable of 120 Hz full frame readouts. For standard imaging, the operating frame rate range for MIRAC-5 is between 0.1 to 85 Hz. The ROIC utilizes a capacitive transimpedance amplifier (CTIA) to capture and readout the signal \citep{jarron2023AN....34430103L} which should preclude persistence \citep{smith2008SPIE.7021E..0JS}. 
    %One useful property of CTIA architecture is that it holds the pixel P-N junctions constant, preventing the undesired filling of charge traps that can induce persistence in detectors \citep{smith2008SPIE.7021E..0JS}. 

    \begin{table}[ht]
        \centering
        \begin{tabular}{c|c|c}
        Parameter     &  Value & Units\\
        \hline
        GeoSnap Format & 1024x1024 & pix \\
        Pixel Pitch & 18 & $\mu$m \\
        Wavelength Range (Half Peak QE Cutoffs)     & 2 - 13 & $\mu$m \\
        Average QE in Range & 65 & \% \\
        Operating Frame Rates & 0.1 - 85 & Hz \\
        Operating Temperatures & $>$ 35 & K \\
        Gain & 83 & e-/ADU \\
        Well-Depth & 1.2E6 & e- \\
        Non-linearity & $<$ 0.1 & \% \\
        Dark Current & $>$ 1.0E4 & e-/s/pix \\
        Instrumental Background & $\sim$ 2.4E5 & e-/s/pix \\
        Read Noise & 133 & e- \\
        Platescale & 0.0192 & arcsec/pix \\
        \hline
        
        \end{tabular}
        \caption{Key properties of the MIRAC-5 GeoSnap. Dark current is measured at operating temperature of 40.5 K and has a bimodal distribution \citep[approximately 4.0E4 e-/s/pix on the left half and 1.0E4 e-/s/pix on the right;][]{jarron2023AN....34430103L}. Non-linearity is $<$ 0.1\% up to full-well after applying a correction. The ``instrumental background'' is measured at operating temperature of 40.5 K and includes undesired emission from various sources within MIRAC-5.}
        \label{tab:properties}
    \end{table}

    The MIRAC-5 GeoSnap's dark current was initially measured at an operating temperature of 40.5 K in the MITTEN cryostat at UM \citep{bowens2020SPIE11447E..37B}. The detector had bimodal dark current of approximately 4.0E4 e-/s/pix on the left half and 1.0E4 e-/s/pix on the right; the limit is thought to be dominated by amplifier glow \citep{jarron2023AN....34430103L}. The MIRAC-5 cryostat suffers from excess background signal which may originate from several sources (ROIC self-emission ``glow'' scattering, thermal emission scattering past the baffle shroud), resulting in an instrumental background on the order of 240,000 e-/s/pix with the two filter wheels set to blank as described in Section \ref{sec:det_measured_perf}. The GeoSnap suffers from 1/f noise which can be mitigated by spatially modulating the signal on the detector through a combination of chopping (via an internal pupil-plane chopping mirror) and nodding. By appropriately selecting a chop/nod frequency, the 1/f noise can be reduced to a tenth or less of the shot noise from the sky, telescope, instrument, and dark for 85 Hz, half-well data.
    
    \subsection{MMT and MAPS AO}

    % AO Status and integration of the two systems
    The MIRAC-5 instrument is used at the MMT Observatory, taking advantage of the 6.5 meter telescope and the (currently undergoing commissioning) MAPS AO system \citep{morzinski2020SPIE11448E..1LM,morzinski202410.1117/12.3019524}. MAPS utilizes a 336-actuator adaptive secondary mirror that can pair with its visible or infrared pyramid wavefront sensors. It operates on a 1-kHz control loop and utilizes only 300 W total power consumption, enabling the use of passive cooling.

    A dichroic is the sole optical element between the MMT secondary and the BLINC optical entrance and is used to split the optical path between MAPS and MIRAC-5. The current dichroic is a temporary option with low transmission (40\% to 50\%). This reduces the final throughput of MIRAC-5 and increases the received background emission. An improved dichroic will be installed for future runs with transmission $>$ 90\% and throughout the paper we provide reference values for the anticipated results with the future dichroic.

    \subsection{Data Acquisition and Modes}
    %Not trying to get to far into the weeds, that's something for a user manual
    %But just enough that a potential user would understand the general process
    %Detector is controlled via gsnap which can be used to take images
    %Other MIRAC-5 features are controlled via an INDI routine (including filter positions, heaters, chopper?)
    %Other systems like MAPS or the telscope are controlled by the AO team and the operator from the same control room
    %A script can be used at the end of operations to update headers with information from outside routines like the telescope info
    %Operational Modes: Staring versus some combo of chop/nodding
    The MIRAC-5 detector is controlled through a custom software package named \texttt{gsnap} which can be used to manipulate detector specific properties including frame rate and initiate exposures. Other features of the MIRAC-5 instrument, including the filter wheels and heaters, are controlled via the Instrument-Neutral Distributed Interface (\texttt{INDI}) software drivers \citep{downey2007indi}. The telescope itself and MAPS AO, are controlled independently. At the end of an observing night, the MIRAC-5 computer retrieves relevant telemetry from a MariaDB database server and populates the headers of observations appropriately. MIRAC-5 can be run in a staring mode or a chop/nod mode. %As the detector suffers from 1/f noise (see Section \ref{sec:det_measured_perf}), it is inefficient to use staring mode for longer exposures or for fainter objects (relative to the background signal level). Instead, a nod or chop-nod sequence should be implemented to ensure the all observations remain shot noise dominated, even for longer exposures (further detailed in Section \ref{sec:programtheory}).

\section{Performance Verification} \label{sec:performance}
    %Then Sky Background and Flatfields
    %Then Throughputs
    %Then Delivered Image Quality (ideal, measured, expected with MAPS)
    %Then Observing Efficiency with staring, nodding, chopping efficiencies
    We describe the performance of MIRAC-5, using data from a commissioning run on May 22nd, 2024 (with a journal of observations given in Table \ref{tab:of_observations}). Data from other runs is used as necessary and we make explicit note for each instance. Our observations took place with MAPS AO providing a static secondary mirror. As such, our image quality depends on the nightly seeing. For all data, we apply a bad pixel mask generated from a set of 128 frames of dark data at the corresponding frame rate. The bad pixel mask excludes data points with signals over 5 median absolute deviations (MAD) from the mean value of all pixels in the 128 frame average. The mask also excludes pixels that exhibit excessive noise (5$\sigma$ from the median standard deviation (STD) of all pixels) or no noise (a STD equal to 0 due to saturated or non-responsive pixels). To capture low or zero response bad pixels, we also utilize the low spatial frequency map of the array described in Section \ref{sec:det_measured_perf}. We divide N-band sky data by this map and then apply the same masking procedure as above. This procedure identifies bad pixels missed using only the dark data. Combined, the two masks typically result in 94.5\% pixel operability.

    \begin{table}[]
        \centering
        \begin{tabular}{c|c|c|c|c|c}
             Target & Time (UTC) & Nod & Band & Frame Rate (Hz) & Frames \\
             \hline
             Alpha Boo & 3:11 & A & K-band & 40 & 1000 \\
             Alpha Boo & 3:12 & B & K-band & 40 & 1000 \\
             Sky & 3:13 & - & K-band & 40 & 1000 \\
             Sky & 3:14 & - & H-band & 40 & 1000 \\
             Alpha Boo & 3:14 & B & H-band & 40 & 1000 \\
             Alpha Boo & 3:15 & A & H-band & 40 & 1000 \\
             Alpha Boo & 3:15 & A & M' & 40 & 1000 \\
             Alpha Boo & 3:16 & B & M' & 40 & 1000 \\
             Sky & 3:17 & - & M' & 40 & 1000 \\
             Sky & 3:17 & - & M-band & 40 & 1000 \\
             Alpha Boo & 3:18 & B & M-band & 40 & 1000 \\
             Alpha Boo & 3:19 & A & M-band & 40 & 1000 \\
             Alpha Boo & 3:21 & A & L' & 40 & 1000 \\
             Alpha Boo & 3:21 & B & L' & 40 & 1000 \\
             Sky & 3:22 & - & L' & 40 & 1000 \\
             Sky & 3:24 & - & W0870 & 40 & 1000 \\
             Alpha Boo & 3:25 & B & W0870 & 40 & 1000 \\
             Alpha Boo & 3:26 & A & W0870 & 40 & 1000 \\
             Alpha Boo & 3:29 & A & Amm. & 40 & 1000 \\
             Alpha Boo & 3:30 & B & Amm. & 40 & 1000 \\
             Sky & 3:31 & - & Amm. & 40 & 1000 \\
             Sky & 3:33 & - & W1252 & 40 & 1000 \\
             Alpha Boo & 3:34 & B & W1252 & 40 & 1000 \\
             Alpha Boo & 3:35 & A & W1252 & 40 & 1000 \\
             Alpha Boo & 3:36 & A & N' & 85 & 2000 \\
             Alpha Boo & 3:37 & B & N' & 85 & 2000 \\
             Sky & 3:38 & - & N' & 85 & 2000 \\
             Sky & 3:41 & - & N-band & 85 & 2000 \\
             Alpha Boo & 3:42 & B & N-band & 85 & 2000 \\
             Alpha Boo & 3:43 & A & N-band & 85 & 2000 \\
             \hline
             Dark & 7:24 & - &K-band + Blank &  40 & 500 \\
             Dark & 7:25 & - &N-band + Blank &  85 & 1000 \\
             Dark & 7:26 & - &Blank + Blank &  85 & 1000 \\
             Dark & 7:27 & - &Blank + Blank &  40 & 500 \\
             \hline
        \end{tabular}
        \caption{A journal of observations taken on May 22nd, 2024.}
        \label{tab:of_observations}
    \end{table}

    \subsection{Detector Performance}
    \label{sec:det_measured_perf}
    %(RN, DC, 1/f, QE variations)
    We recalculate the read noise and dark current of MIRAC-5 from data obtained at the telescope. The ``darks'' for MIRAC-5 include actual dark current from the detector as well as signal originating from the interior of MIRAC-5/BLINC (or navigating through it). To avoid confusion, we henceforth will refer to these as ``instrumental darks''. We utilize data taken on March 19th, 2024 with the detector held at 40.5 K and both filter wheels set to the blank position. To determine the instrumental dark and read noise, we utilize the T\_int feature of our detector which allows us to only expose for a percentage of the total frame time set by the user selected frame rate. We take 50 images per T\_int percentage, stepping from 20\% T\_int to 100\% T\_int in 20\% steps. We combine the frames of the same T\_int using a mean after excluding bad pixels with the mask. Using the gain reported in \cite{jarron2023AN....34430103L} of 83 e-/ADU (re-verified with on-sky data to agreement within 1$\sigma$), we convert the data into e- per pixel and compare to frame integration time. 
    % We then plot these trendlines (with the bias removed) for 5 Hz and 85 Hz detector frequency in Figure \ref{fig:dark_ramp}.
    % \begin{figure}
    %     \centering
    %     \includegraphics[scale=0.5]{figures/m5_dark_ramp_variable_framerates.PDF}
    %     \caption{Ambient dark ramps using the Tint feature of GeoSnap for two different detector frame rates. We fit a line to the data in order to estimate the ambient dark current.}
    %     \label{fig:dark_ramp}
    % \end{figure}
    %all electron and count conversions utilize the 83 e-/ADU calculated in Jarron's paper
    For the 5 Hz darks at 40.5 K, we recover a bias value of 1043 ADU/pix and an ambient dark current of 2.1E5 e-/s/pix. For the 85 Hz darks, we recover a bias value of 1012 ADU/pix and an ambient dark current of 2.7E5 e-/s/pix. Differences between the two bias values and two ambient dark currents for different frame rates has been observed in other GeoSnap detectors \citep{bowens2024arXiv240520440B} although lower dark current was observed at higher frame rates. Laboratory testing at 40.5 K of the MIRAC-5 GeoSnap dark current found a max of 4.0E4 e-/s/pix in the MITTEN cryostat \citep{jarron2023AN....34430103L}.

    To estimate the detector read noise, we perform frame-by-frame subtraction of one thousand 85 Hz 100\%-Tint dark frames taken on May 22nd, 2024, recovering 500 pair subtracted frames. By performing frame-by-frame subtraction, we should eliminate 1/f noise contributions. The expected ambient dark current shot noise for a single 85 Hz frame is 56.4 e-/pix (0.68 ADU/pix). We solve for the temporal variance across each pixel in the 500 pair subtracted frames and determine that the median noise per pixel per frame is 174 e-/pix (2.1 ADU/pix). Assuming the read noise and ambient dark current shot noise add in quadrature to achieve this total noise, the read noise per pixel per single frame is 165 e-/pix (2.0 ADU/pix) compared to 140 e-/pix reported in \cite{jarron2023AN....34430103L}.

    Using the same data set, we estimate the impact of the 1/f noise by performing frame-by-frame subtractions, varying the number of frames read between subtractions. Confirming \cite{jarron2023AN....34430103L}, we find that the 1/f noise is only relevant compared to the half-well shot noise if one does not perform spatial modulation of their source at a sufficient rate (defined relative to the background signal level). Exact values can be checked with the calculator in Section \ref{sec:programuse} but, in general, to reduce the 1/f noise to an acceptable level, high frame rate data such as N' images will require chopping and nodding but low frame rate data such as L' or M' images can be modulated with just nodding.

    % \begin{figure}
    %     \centering
    %     \includegraphics[scale=0.5]{figures/m5_dark_one_over_f.PDF}
    %     \caption{The 1/f trend in the MIRAC-5 GeoSnap for 85 Hz 100\% Tint data. By performing frame-by-frame subtraction with different spacings of frames, one can highlight how the magnitude of the 1/f noise increases with increasing frames between modulations.}
    %     \label{fig:1/f}
    % \end{figure}

    We estimate the pixel-to-pixel QE variations of MIRAC-5's GeoSnap. We begin with a mean N-band image produced from 500 sky frames (see Section \ref{sec:skybackground} for further information on sky analysis). The image shows a gradient across it such that it varies from 6000 ADU/pix on the left side to 8000 ADU/pix on the right side, a result of vignetting within the MIRAC-5 and BLINC optical path. We assume that this trend dominates over low spatial frequency QE variations such that by removing it, all that will remain are pixel-to-pixel QE variations. We utilize the \texttt{scikit} module \texttt{RadiusNeighborsRegressor} to create a 2D interpolation of the sky data. We alter the program such that a pixel's own value is excluded from the interpolation and instead is found via a distance weighted average of all pixels within a 15 pixel radii of the center. 
    %The original mean sky image and the resulting neighbor regressor map are shown in Figure \ref{fig:qe_map}.
    \begin{figure}
        \centering
        \includegraphics[scale=1]{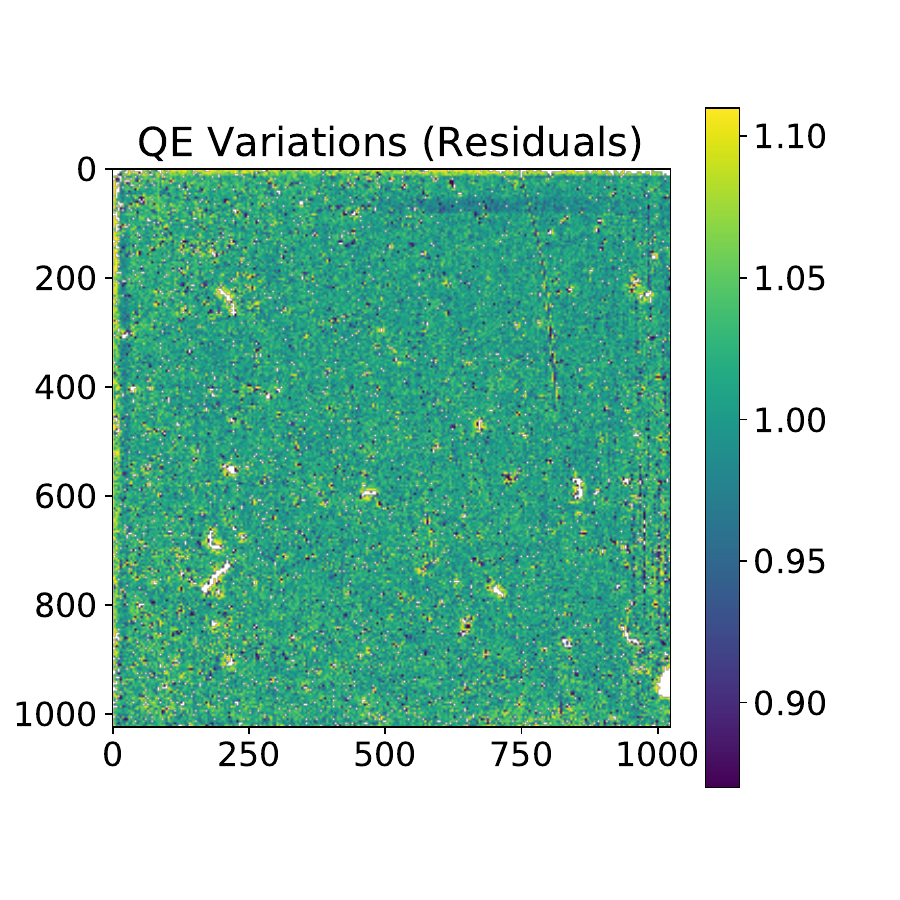}
        \caption{
        %(Top-Left): A mean produced from 500 N-band sky images. Vignetting caused by MIRAC-5 or BLINC causes a gradient on the left side of the image. (Top-Right): A low-spatial frequency behavior map produced by averaging signal of all pixels within 15 pixels of the central pixel, weighted by distance. (Bottom): 
        The mean sky image divided by the low-spatial map with a bad pixel map applied. The variations from unity are the result of pixel-to-pixel variations in QE. Bad pixels are marked in white.
        }
        \label{fig:qe_map}
    \end{figure}
    We divide the mean sky image by the low-spatial frequency regressor map, obtaining a map of QE pixel-to-pixel variations (Figure \ref{fig:qe_map}). We find that the pixel QE histogram is well fit by a Gaussian with mean of 0.990 and sigma of 0.040. This is comparable to 1 - 5 micron MCT detectors such as those in NIRCam with pixel-to-pixel amplitude variations on the order of several percent, particularly NIRCam's A5 detector with 10\% amplitude variations in a crosshatch pattern \citep{schlawin2021AJ....161..115S}.

    % During commissioning, we determined that internal reflection of detector glow off of a filter could increase the noise in the science data. As MIRAC-5 has two filter wheel, we set the nearer filter wheel (relative to the detector) to a desired filter for testing and the farther filter wheel to blank, comparing the measured signal on the detector with the scenario where both wheels are set to blank. For the K-band, we find that the dark current rises to 3665 ADU/s/pix, indicating a small but non-negligible internal reflection of detector glow off of the filters. In N-band, this results in... Compared with other sources of noise, this internal reflection is... 
    %RPB: Probably will be negligible but we'll see once we have N-band. Could not check this in January due to which filters were on which wheels
    
    %For the moment, this section primarily uses the data from January 18th, 2024, taken at 38 K with a slight cloud cover overhead
    %Some data from October 2023 and June 2023 is used to verify K and L throughputs

    % Skies
    %     Calculate emission in sky images and comment on low frequency noise
    %         Also compare with the backgrounds in the images with stars. Do they match?
    %     Determine the variation of the sky throughout the night, compare with the estimated PWV from the paper Jarron found throughout the night
    %         Comment on weather from poor nights in January?
    \subsection{Effective Background and Flats}
    \label{sec:skybackground}
    We characterize the effective background of our images in a range of MIRAC-5 bands using approximately half-well sky frames, taken on May 22nd. Effective background includes contributions from the sky, telescope, and instrument. In Table \ref{tab:skies}, we catalog observations and present effective background levels after subtracting by an equivalent frame rate mean dark made from 1000 (40 Hz) or 2000 frames (85 Hz). We present predicted effective backgrounds following the upgraded dichroic installation for all bandpasses except H- and K-band\footnote{The H and K-band filters are located on the second filter wheel, nearer to GeoSnap. Testing of detector signal with both filter wheels set to blank versus the second wheel set to K revealed excess detector signal on the order of 1.0E4 e-/s/pix. As the observed K-background was 3.0E4 e-/s/pix, this reflection could account for a significant portion of the effective background for these low signal bandpasses.}. These predictions were verified based on preliminary analysis of L' and N' data taken in Nov. 2024 with no dichroic.

    \begin{table}[]
        \centering
        \begin{tabular}{c|c|c|c|c}
            Filter  & Freq. & Frames    & Eff. Back.  &  Future Eff. Back.\\
                    & (Hz)  &           & (e-/pix/s) & (e-/pix/s)\\
                    \hline
                    &&&\\
             H-band      & 40     & 1000        & 0.023E6 & -     \\ %7 ADU/frame/pix
             K-band     & 40    & 1000       &  0.030E6 & -  \\ %9 ADU/frame/pix
             L'   & 40    & 1000       &   0.23E6 & 0.12E6\\ %70 ADU/frame/pix
             M'     & 40    & 1000       &  2.0E6 & 0.77E6\\ %609 ADU/frame/pix
             M-band     & 40    & 1000       & 2.7E6 &1.4E6\\ %824 ADU/frame/pix
             W0870     & 40    & 1000       & 24E6 & 9.9E6\\ %7280 ADU/frame/pix
             N-band    & 85    & 2000       &  84E6  & 34E6\\ %11973 ADU/frame/pix
             Amm.     & 40    & 1000       & 15E6 & 5.6E6\\ %4590 ADU/frame/pix
             N'     & 85    & 2000       &  41E6 & 16E6\\ %5864 ADU/frame/pix
             W1252 & 40 &   1000    &   15E6 & 5.9E6\\ %4585 ADU/frame/pix
        \end{tabular}
        \caption{Effective background in key MIRAC-5 bandpasses. Frames are combined into a mean frame before being dark subtracted. Final effective background is found from a median of the mean frame. We model the expected future effective background following the dichroic replacement in all bands except H and K. This calculation accounts for both the improvement in dichroic throughput and the reduction in dichroic emission (see Section \ref{sec:programuse}).}
        \label{tab:skies}
    \end{table}

    We calculate the variance in sky flats across a night utilizing two sets of 85 Hz N-band sky images taken four hours apart on May 19th, 2024. We limit our data to the central 200x200 pixel region and then ratio the flats. After applying a bad pixel mask, we find that the ratio of the two flats are fit by a Gaussian with a mean of 0.9959 and a sigma of 0.0060, representing a Signal-to-Noise Ratio (SNR) on the order of 166. The variations between the two flats may be a result of slight shifts in the pixel level dark current throughout the night and will be further assessed in the future.

    We also calculate the changes in the effective background signal across a night. Using two more sets of 85 Hz N-band sky data from May 19th, 2024, we extend the baseline to a 6 hour period, and we determine that the effective background levels changed at only about a 2.5\% level across the observing window. We utilize data from National Ocean and Atmospheric Administration's Geostationary Operational Environmental Satellites (GOES) in conjunction with the python package \texttt{fydor} to estimate precipitable water vapor at zenith the MMT during the night \citep{meier2021A&A...649A.132M}. GOES-16 provides weather data in 10 minute intervals across much of the North America continent. We find changes on the order of 2 mm across a night are typical. We utilize ESO's \texttt{SkyCalc} which was developed by a team at the Institute for Astro- and Particle Physics at the University of Innsbruck based on the Cerro Paranal Sky Model \citep{noll2012A&A...543A..92N, jones2013A&A...560A..91J} to estimate the impact of these variations in water vapor on the expected sky background. In \texttt{SkyCalc}, we vary the water vapor from 4.5 to 2.5 mm (matching estimated values on May 19th), finding it can decrease N-band emission by approximately 33\%. As N-band sky emission is expected to display significantly greater variance over the course of a night, we conclude that our effective background must be dominated by telescope and instrument contributions. This is reinforced by the temperature logs which showed that the temperature of the MMT changed by approximately 1 K during the night. Feeding a 1 K change into our exposure time calculator described in Section \ref{sec:program}, we confirm that the expected change to the telescope/instrument emission is 2\% and that the telescope/instrument contribution dominates over the sky contribution. This will remain true even after we reduce the telescope contribution by implementing the replacement dichroic.

\begin{figure}
        \centering
        \includegraphics[width=0.5\linewidth]{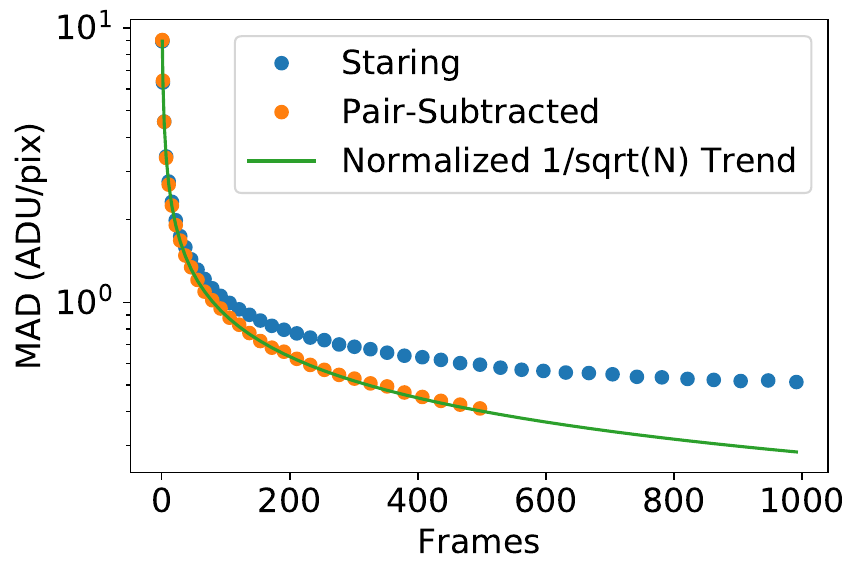}
        \caption{The scaling of the spatial noise estimated via MAD in mock staring mode and chop-nod mode N' sky images. Staring mode images diverge from the 1/sqrt(N) noise scaling due to the non-Poisson 1/f noise floor.}
        \label{fig:noise_pair}
    \end{figure}

    We estimate the scaling of the noise with time in a series of sky frames. We utilize 1000 frames of N' sky data taken on May 22nd  and focus on a 100x100 pixel subregion. No dark subtraction or flat fielding is applied. We do apply a bad pixel mask and then for each pixel estimate the temporal standard deviation (STD) across some number of frames. We find a median temporal STD of 8.9 ADU/pix. To test how the noise scales with number of frames, we average together $\lbrace 4, 16, 32 \rbrace$ adjacent frames. We find that the median temporal STD drops to 5.0 ADU/pix, 2.8 ADU/pix, and then 2.0 ADU/pix, respectively. This represents close to 1/sqrt(N) scaling, as would be expected for a Poisson noise dominated observation. The discrepancy from 1/sqrt(N) then represents the impact of 1/f noise. To show this, we reduce the data by applying dark subtraction and a flat field made from independent N' sky data. We can then calculate the noise in a spatial region via the median absolute deviation (MAD). We input 500 frames to represent staring mode and plot their spatial noise per pixel as well as the 1/sqrt(N) trendline in Figure \ref{fig:noise_pair}. We also produce 500 pair-subtracted frames by performing adjacent subtractions (frame 1 by frame 2, frame 3 by frame 4, etc.). This process effectively eliminates 1/f contributions. We divide the resulting noise by sqrt(2) to put it into the context of a single frame and then plot it on the same figure. As seen in Figure \ref{fig:noise_pair}, there is an approximately 0.2 ADU/pix difference between the final staring mode and the ideal trendline, implying a systematic noise threshold of about 0.41 ADU/pix. This systematic noise floor is dominated by 1/f noise but may contain other elements such as a systematic noise introduced by the mean dark frame subtraction (which would cancel out in pair-by-pair subtraction). 
    
    The SNR for flats (made from skies and darks) and darks will impact the final uncertainty in a reduced science image. In terms of SNR for typical frames, a sky frame will have on the order of 6,000 ADU/pix or 5E5 e-/pix. With a shot noise of $\sim$22 e-/pix after averaging 1000 frames and a 1/f noise threshold of 14 e-/pix, the sky data will have a SNR on the order of 19,000 (though as noted, variations of flats across the night imply a lower SNR). Dark data may be a limiting factor for a reduced image's final SNR. An 85 Hz dark frame will result in about 3200 e-/pix as derived in Section \ref{sec:det_measured_perf}. With a shot noise of about 1.8 e-/pix after averaging 1000 frames and a 1/f noise threshold of 14 e-/pix, the dark data will have a SNR on the order of approximately 225 and thus will contribute some noise to all reduced images (and flats).
    
\begin{figure}
        \centering
        \includegraphics[width=0.5\linewidth]{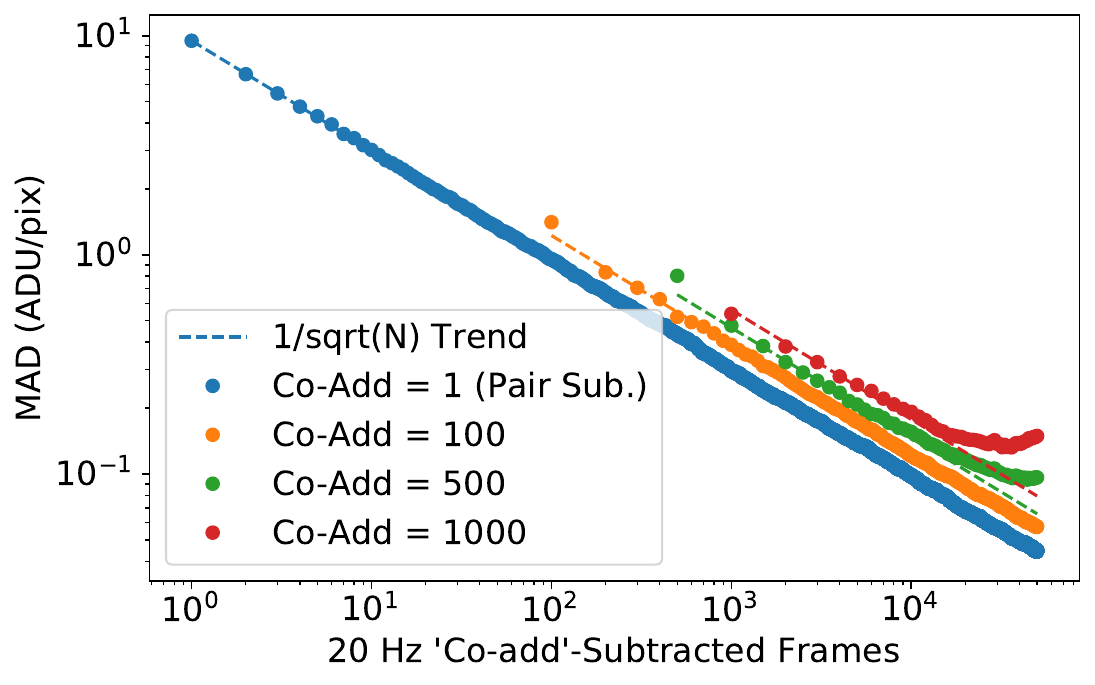}
        \caption{Similar plot to Figure \ref{fig:noise_pair} but in a log-log scale and calculated with a longer time baseline (83 minutes) dataset taken during the November 2024 with a flat secondary. We once again show pair-by-pair frame subtraction but have also added curves for the scaling if the data is temporally co-added before subtraction. The co-added data shows higher noise at all points and long time baseline co-adding causes an early departure from the 1/sqrt(N) scaling.}
        \label{fig:noise_pair_long}
    \end{figure}

    To test scaling for longer observations, we use N' sky data taken on Nov. 20th, 2024 with the secondary held flat. These data, taken at 20 Hz for 100,000 frames, represents about 83 minutes of exposure time and continues to show the noise following 1/sqrt(N) scaling (Figure \ref{fig:noise_pair_long}). We also demonstrate the impact of temporally co-adding data before pair-wise subtraction, finding that not resolving time-variable fluctuations in the background on a per-frame and per-pixel bases causes a significant increase in noise (which is greater for more co-adding).

    % The large signal in the skies represents one of the greatest sources of noise in mid-IR observations. It is a combination of atmospheric emission and telescope emission, with the latter dominating based on our estimates in Section \ref{sec:program}.
    % %RPB: Should we comment on disentangling these factors? As far as I am aware, we cannot do that
        
    % Throughputs and Backgrounds
    %     Details on observations of calibration stars in each filter. Provide information on how the aperture photometry was carried out, the resulting SNR, and the estimated throughput for the filter
    %     With throughputs in hand, provide estimated background emissions as well
    \subsection{Throughputs}
    \label{sec:throughputs}
    We calculate the throughput of MIRAC-5 in all major bands utilizing images of Alpha Boo (Arcturus) taken on May 22nd, 2024. Alpha Boo is a K1.5 Giant \citep{keenan1989ApJS...71..245K}. We draw Vega magnitudes from a spectral model produced for Alpha Boo by \cite{cohen1995AJ....110..275C}: H-mag of -2.96, K-mag of -3.04, L-mag of -3.15, M-mag of -2.93, 8.7-mag of -3.12, N-mag of -3.14, and 11.7-mag of -3.16. These magnitudes have uncertainties of 0.01 to 0.03. 
    %We cross-check these values with the 2MASS \citep{cutri2003yCat.2246....0C} K-mag of -2.91 $\pm$ 0.17. For longer infrared magnitudes, we apply conversions from the measured K-mag based on the work of \cite{pecaut2013ApJS..208....9P}, substituting their WISE filter conversions, W1, W2, and W3, for our L', M', and N' filters, respectively. We recover values of L' = -2.94 $\pm$ 0.17, M' = -2.88 $\pm$ 0.17, and N' = -2.90 $\pm$ 0.17. Given the uncertainty on the 2MASS K-mag, we conclude the results are consistent and utilize the \cite{cohen1995AJ....110..275C} magnitudes.

    We observe Alpha Boo in each bandpass for 1000 frames (for 40 Hz observations) or 2000 frames (for 85 Hz observations) at each nod position. Observations are performed with a flat secondary. We perform an approximately 5 arcsecond nod for each observation, doubling the total data for analysis. We then move off target and take 1000 to 2000 sky frames for use in flat fielding. Finally, we took dark measurements at 40 Hz and 85 Hz with 1000 and 2000 frames, respectively. See Table \ref{tab:of_observations} for details.

    To perform automated photometry, we must prepare our nod A and B frames. First, we construct appropriate flats for the data. We then create mean frames for nod A and B data after flat fielding. These two mean frames are used for nod-subtraction purposes, removing background signal to allow for automated stellar centroiding and registration. For each raw nod A frame, we perform flat-division before subtracting them by the mean nod B frame. This produces a frame with a positive image of the star (from one frame) and a negative image of the star (from a mean frame). Only the positive star will be used in future alignment. We repeat the process for the nod B data. To see the individual nod A frames following the mean B subtraction, please refer to the mosaics presented in Appendix \ref{ap:singleframe}.

    Using \textit{PYNPOINT} \citep{amara2012MNRAS.427..948A,stolker2019A&A...621A..59S}, we input the frames and then perform a bad pixel smoothing by iterating three times and replacing outliers that deviate from their neighbors by more than 3-sigma with a mean of the neighbor values. We crop a region around the positive image of the star of 8.4 arcsecond width and fit Gaussians in X and Y to the PSF. To remove poorly fit frames, we reject any frames with values 4-MAD below the mean signal value within a 0.285 arcsecond radius. On average this removes 1\% of frames. The fits are then used to align the nods frame by frame utilizing a fifth order spline at super resolution. We produce mean frames for position A and B, and we then confirm that the two return identical FWHM values. We average the two mean frames together (with the star centrally aligned in both), producing an overall mean frame for the nod-pair. 
    
    With the mean images prepared for each filter, we generate radial profiles and encircled energy profiles using the \texttt{astropy} circular aperture photometry function \citep{astropy2022ApJ...935..167A}. We optimize the central aperture's radii by stepping outwards in 1 pixel steps starting from 10 pixels, calculating the aperture photometry at each step using an outer annuli of 5 pixels width starting 15 pixels beyond our current central radii. We continue stepping outward until the increase in stellar signal in electrons is smaller than the shot noise in the central aperture background (as estimated from the RMS in the sky aperture). To avoid overlap with the negative image of the star, we utilize a max aperture central radius of 175 pixels (3.33 arcseconds).

\begin{figure}
        \centering
        \includegraphics[width=0.5\linewidth]{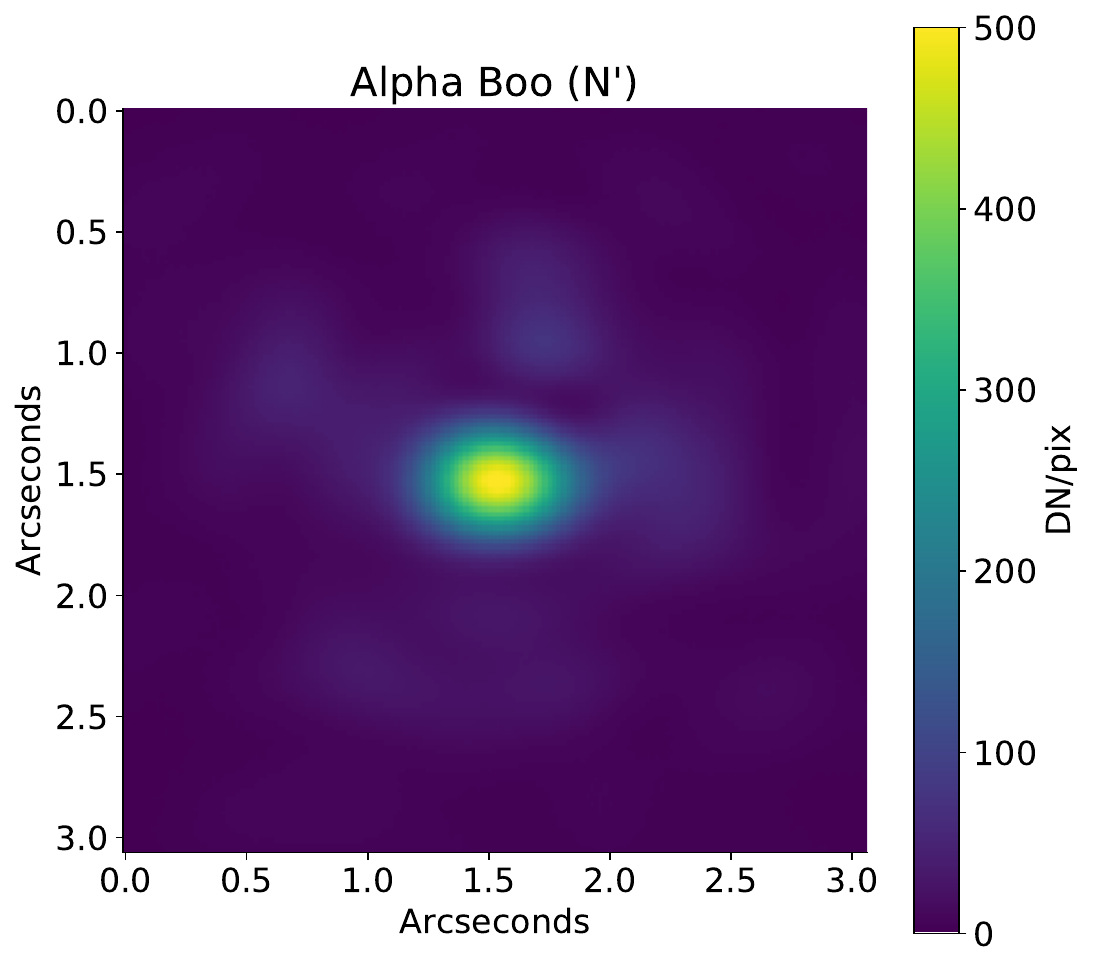}
        \caption{The final aligned imaged for Alpha Boo after combining 2000 A nod and 2000 B nod frames using \textit{PYNPOINT} \citep{amara2012MNRAS.427..948A,stolker2019A&A...621A..59S}. The PSF for this particular image is given in Figure \ref{fig:imqual}. The PSF is approximately 25\% longer in the x-direction than y-direction which may be a result of slight pupil misalignment during the May run coupled with a section of actuators on the secondary mirror's edge that were out of their nominal flat position.}
        \label{fig:alphaboonprime}
    \end{figure}

    We lock in the central aperture radii and then optimize the outer annulus. Starting from 15 pixels beyond the edge of the central radii, we increase the outer annulus width by 3 pixels per step. At each step, we calculate the standard deviation of the pixels within the outer annulus. Once the new standard deviation exceeds the prior one (indicating that we are introducing significant residual background variations into the photometry by continuing to expand the annulus), we truncate the loop and use the prior annulus width. We define a max outer annuli of 215 pixels (4.09 arcseconds). In Table \ref{tab:throughputs}, we compare the observed photometric values of Alpha Boo with the known apparent magnitudes and then calculate the total throughput per bandpass for MIRAC-5. We plot the final aligned image for an N' image of Alpha Boo in Figure \ref{fig:alphaboonprime}.

    \begin{table}[h!]
        % \centering
        \begin{tabular}{c|c|c|c|c|c|c|c}
        Band    &Frame                &Sky    &Source     &Known      &Measured to  &T$_{total}$ &T$_{total}$\\
                &Rate              &Median &Signal  &App. Mag.  & Expected Ratio & & (Future)\\
            &(Hz)     &(ADU/pix)    & (ADU)     &       &&(\%)&(\%)\\
        \hline
         H-band   &40            & 7.0     & 3.96E5     & -2.96& 0.0014&0.1 & 0.001\\        
         \hline
         K-band   &40             & 9.0     & 7.03E6     & -3.04& 0.28&3.3 & 7.6\\        
         \hline
         L'   &40             & 70     & 1.26E7     & -3.15 & 0.39&13&22\\
        \hline
        M'   &40             & 610     & 2.18E6     & -2.93& 0.34&  5.5&16\\
        \hline
        M-band   &40            & 820     & 4.98E6     & -2.93& 0.67&  12&39\\
        \hline
        W0870   &40          & 7300     & 1.78E6     & -3.12& 0.30&  9.7 & 20\\
        \hline
        N-band   &85       & 12000     & 1.59E6     & -3.14& 0.23&  7.2 & 15\\
        \hline
        Amm.   &40        & 4600     & 5.47E5     & -3.14& 0.27&  8.8 & 17\\
        \hline
        N'   &85       & 5900     & 6.15E5     & -3.14& 0.30 & 8.0 & 17\\
        \hline
        W1252   &40         & 4600     & 3.15E5     & -3.16& 0.29&  4.8&11\\

        \hline
        \hline
        \end{tabular}
        \caption{Aperture photometry results for Alpha Boo on May 22nd, 2024. Sky medians are found from the median pixel value in a dark subtracted mean image of nod position A. The measured to expected ratio is a comparison the measured source intensity versus an expected source signal converted from the known apparent magnitude. T$_{total}$ gives the total transmission (including QE losses), both for the current and future dichroic. Data taken in L' and N' on Beta Gem, Beta Peg, and Beta And in Nov. 2024 with a flat secondary and no dichroic returned throughputs of 19.5\% and 12.3\%, respectively. Discrepancies from the predicted future transmission may be due to errors in the aperture photometry given the very low image quality during that run without AO support.}
        \label{tab:throughputs}
    \end{table}

    %  Image Quality
    %     Comment on the current state of image quality given the AO performance
    %     What is our Strehl? Can Oli calculate via Zeemax an ideal PSF (perhaps with some error assumed for MAPS) and then we compare that result with our PSF to get the Strehl?
    \subsection{Delivered Image Quality}
    \label{sec:imagequal}
\begin{figure}
        \centering
        \includegraphics[width=0.5\linewidth]{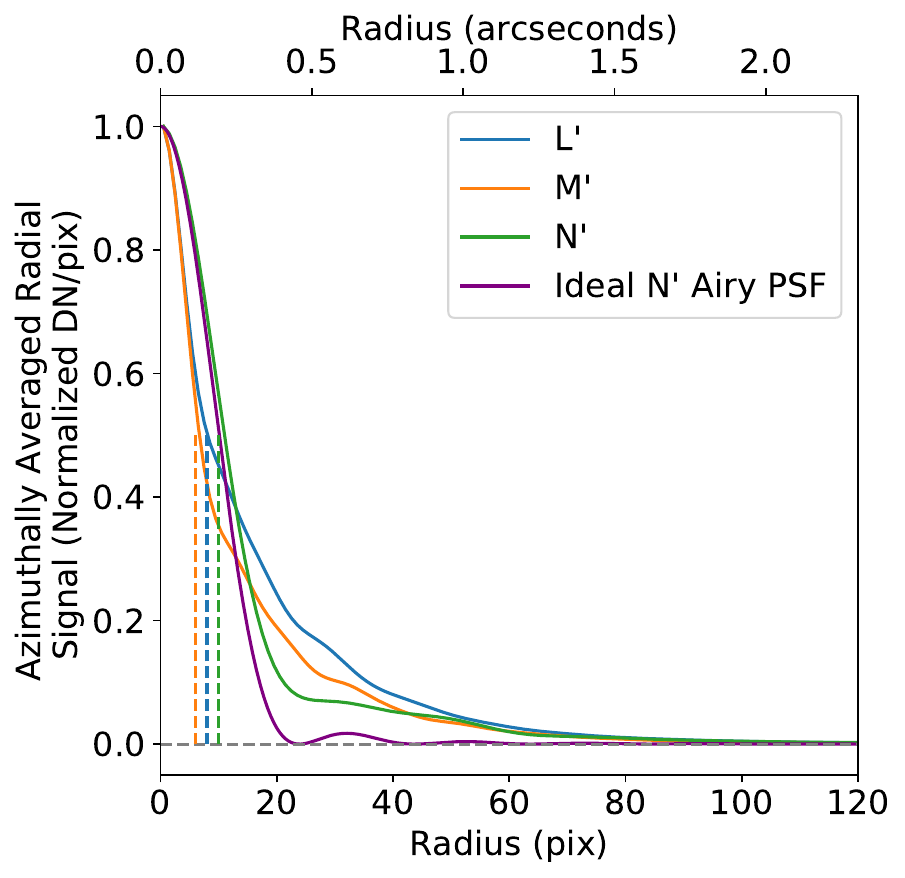}
        \caption{Radial profiles for the mean-subtracted images of Alpha Boo in the L', M', and N' filters. The locations of the FWHM are marked with dashed lines. An ideal Airy profile for the central N' wavelength is shown in purple.}
        \label{fig:imqual}
    \end{figure}

\begin{figure}
        \centering
        \includegraphics[width=0.5\linewidth]{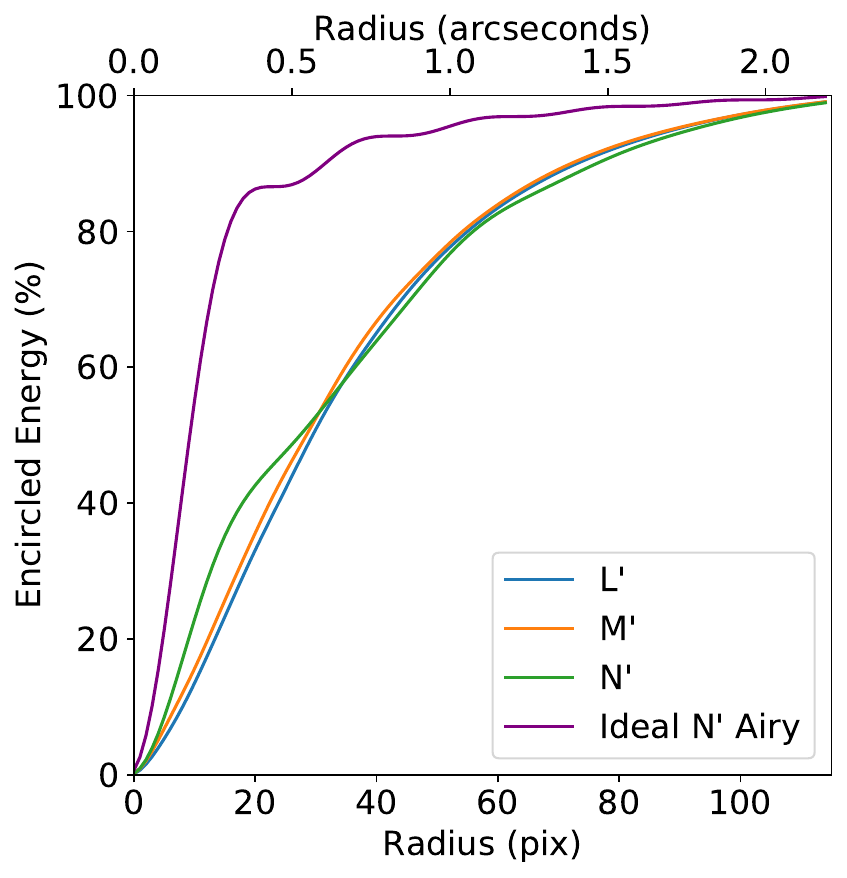}
        \caption{Encircled energy profiles for the mean-subtracted images of Alpha Boo in the L', M', and N' filters. An ideal Airy profile encircled energy for the central N' wavelength is shown in purple.}
        \label{fig:encircled}
    \end{figure}

    %Ideal, then measured, then expected
    Utilizing the mean-subtracted images, we calculate the delivered image quality for MIRAC-5 for these observations. We plot the radial profiles for L', M', and N' positive images in Figure \ref{fig:imqual} and the encircled energies in Figure \ref{fig:encircled}, both over a radius of 120 pixels (2.28 arcseconds). 

    We find a FWHM of 16 pixels (0.31"), 12 pixels (0.23"), and 20 pixels (0.38") in L', M', and N' bands, respectively. For a diffraction limited 6.5 m telescope with an Airy profile, we can estimate the FWHM as approximately 1.02 times the diffraction limit ($\lambda/D$), i.e., 1.02$\lambda/D$. This returns 0.12" in L', 0.15" in M', and 0.37" in N'. We show in  Figure \ref{fig:imqual} that all three bands have diffraction limited cores but that their wings are significantly extended due to atmospheric impact without AO correction. These wings will be reduced once the AO system is operational at which point we will assess if any remaining deviations from the ideal image quality are the result of MIRAC-5/telescope optics or atmospheric turbulence. Since our current PSFs have extended wings compared to the ideal Airy pattern, they encircle less energy at a given radius. The ideal Airy profile encircles 48.3\% of the total energy within its FWHM but we find that we only encircle 8.3\%, 6.8\%, and 19.9\% in the FWHM of L', M', and N', respectively. The N' band currently has a Strehl ratio of 0.75 based on a comparison of the current PSF with an ideal Airy profile containing an equivalent encircled energy. In Section \ref{sec:perfomance_limits}, we utilize both the ideal and current PSFs to estimate the background limited sensitivity for MIRAC-5.

    \subsection{Observing Efficiency}
    %Staring, nodding, and chopping efficiencies
    MIRAC-5's observing efficiency is dependent on the selected operation mode. Staring mode results in the most on-sky observing time, as the only limiting factor once the AO closes loop on a target is the minuscule frame reset time. However, as detailed in \cite{jarron2023AN....34430103L}, GeoSnap suffers from 1/f noise which will dominate over other noise sources in most staring mode applications. Staring mode will thus be inefficient (in terms of realized SNR) for longer exposures.

    To mitigate 1/f noise, MIRAC-5's nodding and/or chopping mode is employed. The MIRAC-5 chopper spatially translates the target approximately 410 pixels or 7.4". MIRAC-5's 1/f noise actually scales by number of frames, not time between exposures \citep{jarron2023AN....34430103L, bowens2024arXiv240520440B}. By subtracting frames in close numeric proximity, users can lower 1/f noise contribution well below the expected shot noise from the effective background. The optimal nodding/chopping frequency is then a function of the frame rate and effective background. This will define an efficiency for each frame rate. For an 85 Hz half-well observation given the currently measured background level (e.g., N-band), it should be sufficient to chop every 5 frames (background shot noise will be on the order of x10 the 1/f noise). This would correspond to a chopper frequency of 8.5 Hz. As the MIRAC-5 chopper is driven by a voice-coil actuator, it goes through a sequence of acceleration, deceleration, and settling each swing. This process takes about 30 ms and a fully cycle will require two swings, costing the user 60 ms. The ideal chopping frequency is a balance of the 1/f noise contribution and the efficiency loss. A full description of this relationship is provided in Section \ref{sec:programtheory}.

    % To minimize the impact of 1/f noise in our measurements, we perform chop-nod subtraction on a subset of data taking for the calibration star X. Star X is first observed in staring mode configuration for X minutes and the resulting data is reduced as described in prior sections. We then observe star X with the internal BLINC chopper set to a frequency of X Hz, resulting in a Y throw across the detector FoV. Furthermore, we nod... The resulting data is reduced as follows...
    % %RPB: For chop-nod reduction, would one still do dark subtraction? Technically it should not matter since it will just get subtracted out by the A minus B process.

    % We plot the SNR for the star in the staring mode and the chop-nod mode in Figure B. Along the x-axis, we increase the number of frames used to generate the master frame. We also plot the 1\/sqrt(N) to show the ideal scaling for combining frames on the SNR. As expected from the results in \citet{jarron2023AN....34430103L}, the chop-nod sequence scales as 1\/sqrt(N) at a ratio some fixed value higher than the ideal. The value of this ratio is defined by the chopping frequency selected.

    % One must take care in the appropriate selection of their chop-nod frequency. Although a higher chop frequency will reduce the impact of the 1/f noise, it will also reduce the efficiency of the observation (as more frames will be lost during chops).
    % %RPB: My program should provide users a means to input different chopping rates and see the overall impact on exposure time

\section{MIRAC-5 SNR, Exposure Time, and Calculator} 
\label{sec:program}
\subsection{SNR and Exposure Time}
\label{sec:programtheory}
    The MIRAC-5 SNR for a single frame can be defined as follows:

    \begin{equation}
    SNR_o = \frac{St}{\sqrt{St + n((A + T_{ti} + D)t + \sigma_{RN}^2 + \sigma_{1/f}^2}}
    \end{equation}
    
    In the above equation, $S$ represents the signal from the source in e-/s in an area of $n$ pixels, $t$ the detector integration time in seconds for a frame, $A$ the atmospheric flux in e-/s/pix, $T_{ti}$ the telescope and instrument flux in e-/s/pix, $D$ the dark current in e-/s/pix, $\sigma_{RN}$ the read noise in e-/pix, and $\sigma_{1/f}$ the 1/f noise for a given sampling frequency in e-/pix. $A$ and $T_{ti}$ are challenging to disentangle and are quoted as a single value, the effective background. The 1/f noise follows an empirical relationship:

    \begin{equation}
    \sigma_{1/f}^2 = g^2k_f \left(\frac{\nu_{chopper}}{\nu_{detector}}\right)^\alpha,
    \end{equation}

    \noindent $g$ represents the gain in e-/ADU for the detector, $\nu_{chopper}$ the frequency of the chopper in Hz, $\nu_{detector}$ the frame rate of the detector in Hz, and $k_f$ and $\alpha$ two empirically derived values for the 1/f power law with values of 0.012 (ADU/pix)$^2$ and -1.348, respectively. These values were derived with 85 Hz dark data taken on the mountain and match values found via lab data in \cite{jarron2023AN....34430103L}. Together, $\frac{\nu_{chopper}}{\nu_{detector}}$ is referred to as the ``frequency per frame.''

    As $T_{ti}$ will dominate the noise terms in most use cases, the $SNR_0$ is shot noise dominated. Therefore, the number of science frames, $N_{science}$, required to achieve a desired $SNR_{goal}$ can be written as:

    \begin{equation}
    N_{science} = \left(\frac{SNR_{goal}}{SNR_0}\right)^2
    \end{equation}

    The above relationship will remain true as long as the 1/f noise contribution is insignificant compared to the shot noise, as demonstrated in Figure \ref{fig:noise_pair_long}. MIRAC-5 observations are not perfectly efficient though, as swing and settling times (for chops and nods) create overheads. \textbf{In the following set of equations, our goal is to determine the number of frames for various user-defined options ($\nu_{detector}$, $\nu_{chopper}$, $\nu_{nodding}$) that must be taken in order for the number of usable science frames to reach the required level for a desired SNR.} Not all frames will be usable science frames as some will be imaged during a chop or nod settling time. We begin by estimating the impact of the chopper. A full period of the chopper is comprised of two approximately 30 ms intervals. We define the number of frames lost per chop cycle, $n_{mid\_chop}$ as:
    
    \begin{equation}
    n_{mid\_chop} = 2(\left\lceil t_{swing}*\nu_{detector}\right\rceil+1)
    \end{equation}

    % $$n_{mid\_chop} = 2(\left\lceil t_{swing}*\nu_{detector}\right\rceil+1)$$
    
    In the above, $t_{swing}$ = 0.03 seconds and $\left\lceil \right\rceil$ represents the ceiling function, i.e., rounding the enclosed up to the nearest integer. First, we calculate the minimum number of frames a chop swing will occupy, rounded up as any frame with chopping motion will typically be discarded. We add ``$+1$'' to the number of frames lost because the chop movement is not yet configured to trigger at the frame boundary, therefore costing us two frames during a swing. Finally, we multiply by 2 as a full chop cycle includes two chops. 
    %For a 50 Hz frame rate, we would then have $n_{mid\_chop} = 2(RoundUp(0.03*50)+1) = 2(2+1) = 6$ frames considered unusable.
    
    To determine the total amount of observing time required for a specific SNR, we calculate how many science frames we get across a full cycle of the chopper ($n_{science}$). We define the total number of frames taken during a chop cycle, $n_{total}$ and then solve for $n_{science}$:

    \begin{equation}
    n_{total} = 2*\frac{\nu_{detector}}{\nu_{chopper}}
    \end{equation}

    \begin{equation}
    n_{science} = n_{total} - n_{mid\_chop}
    \end{equation}

    We implement the nodding impact in a similar manner. Nodding is far less frequent than chopping but will take more time to complete. We define the time for a nod to occur as $t_{nodswing}$, i.e., the total time to move from position A to B and reestablish imaging. Nodding will be performed in an A-B-B-A order such that only one nod swing is required per nod cycle (in contrast to the 2 for chopping). Nodding will occur at some frequency, $\nu_{nodding}$, such that the number of frames lost per nod cycle is:

    \begin{equation}
    n_{mid\_nod} = (\left\lceil t_{nodswing}*\nu_{detector}\right\rceil+1)
    \end{equation}

    To simplify our calculations, we calculate the average number of frames within a chop cycle that would be lost because of nodding. Then $n_{science}$ becomes:

    \begin{equation}
    n_{science} = n_{total} - n_{mid\_chop} - n_{mid\_nod}*\frac{\nu_{nodding}}{\nu_{chopper}}
    \end{equation}

    The observing efficiency is then:

    \begin{equation}
    e_{obs} = \frac{n_{science}}{n_{total}}
    \end{equation}

    The required ``science'' observing time is:

    \begin{equation}
    t_{science} = \frac{N_{science}}{\nu_{detector}}
    \end{equation}

    And the total observing time (accounting for losses due to chopping and nodding) is:

    \begin{equation}
    t_{obs} = \frac{t_{science}}{e_{obs}}
    \end{equation}
    
\begin{figure}
        \centering
        \includegraphics[width=0.5\linewidth]{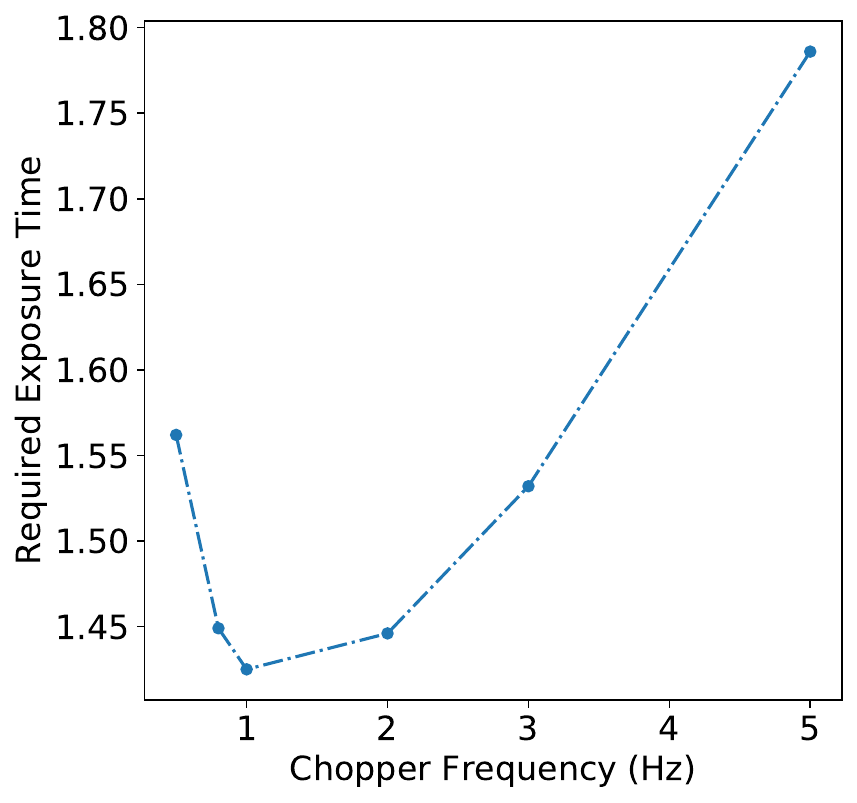}
        \caption{Required exposure time for a theoretical N' target observed at 50 Hz frame rate with background at half-well depth given different chopper frequencies. In this scenario, a 1 Hz chopper frequency results in the lowest total observing time (1.43 hours) or approximately 10\% higher than the ideal exposure time with no 1/f or chopping efficiency losses. Although 1/f noise contributions decrease at higher chopper frequencies, the observing efficiency also declines. Users will need to optimize the chopper frequency to balance those two factors.}
        \label{fig:f_efficiency}
    \end{figure}

    In Figure \ref{fig:f_efficiency}, we demonstrate the total required observing time for a simulated half-well, 50 Hz observation with a N'-mag = 11 target. Although a high chopper frequency eliminates 1/f noise contributions, it greatly reduces the observational efficiency as well. A chopper frequency at 1 Hz has more 1/f noise but the improved observational efficiency makes up for the noise, resulting in the lowest total observing time to the desired SNR. Users should utilize the program in Section \ref{sec:programuse} to determine the best chop-nod combo and frequency given their target's flux and expected image quality. For bandpasses that require lower frame rates such as L' or M', it is sufficient to modulate only via nodding and therefore avoid unnecessary changes to the optical path induced by chopping.

    % The code described in Section \ref{sec:programuse} can be used to calculate required observing time for a specific SNR. It is recommended that users perform nodding at a frequency of XX. The ideal chopping frequency is more complex and will depend on the use case. Although chopping at 0.1 frequency (per frame) effectively eliminates 1/f noise, the excess time loss due to the decrease in observing efficiency can nullify gains. In Figure \ref{fig:f_efficiency}, we plot an example of this scenario for a simulated half-well, 50 Hz observation with a Np-mag = 11 target. Although chopping at 5 Hz (0.1 frequency (per frame)) results in an extremely low 1/f of 0.97 (ADU/pix)$^2$, the loss in observing efficiency relative makes this chopper frequency a poor choice compared to chopping at 1 Hz (which results in 8.5 (ADU/pix)$^2$ 1/f noise). \textbf{As long as the final SNR does not require a noise level below 8.5 ADU/pix, the 1 Hz chopper frequency will be the correct choice in this scenario.} Generally the ideal chopping frequency is between 0.1 to 0.02 frequency (per frame). Users should utilize the program in Section \ref{sec:programuse} to determine the best chopping frequency given their target's flux and expected image quality.

    In a staring mode configuration, the noise approaches a systematic uncertainty threshold of approximately 0.4 ADU/pix, although the exact value may be higher if low SNR darks or flats are applied. The threshold is reached within approximately 1000 frames, as seen via the blue curve in Figure \ref{fig:noise_pair}. Therefore, the limit on the SNR for staring mode will always be the target signal per frame divided by the systematic noise threshold. Staring mode is ineffective for longer integrations, i.e., fainter targets.

\subsection{Calculator}
\label{sec:programuse}

    %Michael: Include real science results
    %Include proposed science cases

    % Describe the M5 helper program used for SNR estimates
    %     Provide a link to a user friendly version of the program that readers may utilize for M5 runs
        
    % Note improvements to the program after the commissioning run (particularly comment on how assumptions of the background may have changed)

    We provide a \texttt{Jupyter notebook} written in \texttt{Python} that can be used to estimate the required exposure time to achieve a specified SNR. Instructions for the usage of the \texttt{Jupyter notebook} are provided within but a brief description of the code is given here. The program requires several user inputs (bandpass of interest, target properties, frame rate and chopper frequency, etc.) as well as several ``static inputs'' (telescope diameter, f-number, etc.) which are not expected to change.

    The program estimates the radiance of the target assuming a blackbody model or a user supplied spectrum. To simulate source transmission through the atmosphere and mid-IR emission from the atmosphere, the program utilizes data produced by ESO's \texttt{SkyCalc}. We provide several example emission and transmission profiles for different precipitable water vapor levels. We utilize the Cerro Paranal Observatory altitude (2640 m) as it is the closest option to MMT (at 2616 m). We also perform our calculations for a target at zenith. All other parameters (such as scattered starlight and moonlight, moon separation, airglow, etc.) are left at their default values. With the exceptions of molecular emission from the lower atmosphere or close proximity to the moon, these parameters have negligible impacts beyond 4 microns on the emission. Instrumental thermal emission is left at zero as we implement it independently below. From \texttt{SkyCalc}, we obtain radiance and transmission over our detector's range. Atmospheric emission is expected to be small compared to the telescope's emission and therefore slight differences between the \texttt{SkyCalc} parameters and real observing conditions should not be a concern. If users wish though, they can generate their own files from \texttt{SkyCalc} specific to their observation and utilize those.
    
    After sky emission and transmission, the emission from the telescope plus instrument is estimated for a given telescope temperature (specified by the user, default of 280 K) and emissivity (assumed to be 0.1). We assume the emission primarily originates from the primary mirror, secondary mirror, and dichroic. We estimate their emission as a blackbody of the given temperature at a set solid angle such that, given the same emissivity, they should have equal contributions to the total background (assuming that their physical size is larger than the beam width at their location). The current dichroic in MIRAC-5 is not ideal, featuring an emissivity of approximately 0.5, and thus dominating the telescope plus instrument contribution. In future runs, it will be replaced with a low emissivity dichroic: reducing the background and improving the throughput. Accounting for these three surfaces and the per bandpass emissivity on the dichroic, we find that the program's predicted background is between x0.5 to x1.0 the observed background in most filters. Exceptions are found in the H-, K-, and M-bands which may occur because of detector glow reflection (H and K) or highly variable regions of the dichroic's transmission (M). An empirical correction to the telescope plus instrument background is applied per bandpass to match the program to observations.
    
    The program also applies an empirical adjustment to the throughput for the selected bandpass, impacting target, sky, and telescope/instrument light. The user must define the diameter around the target's center to utilize in the SNR calculation. Ideally, a user selects a diameter where beyond it the increase in noise per pixel outweighs the increase in signal within the SNR, typically around the FWHM for our measured encircled energy profiles. The user must also select the PSF shape using either the calculated per band PSF (and corresponding encircled energy) or using the ideal Airy PSF and encircled energy.

    In summary, all three sources are calculated as follows:
    \begin{enumerate}
        \item Target: Flux is calculated for target distance, atmospheric transmission based on provided \texttt{SkyCalc} file, and empirical throughput. The user provides the diameter of interest and the corresponding encircled energy curve (either ideal Airy PSF or the measured PSF for the band).
        \item Sky: Flux is calculated for empirical throughput and atmospheric emission based on provided \texttt{SkyCalc} file.
        \item Telescope plus Instrument: Flux is calculated for empirical throughput and an assumed telescope temperature. An empirical bandpass dependent background adjustment is applied at the end.
    \end{enumerate}

    Instrumental background is measured at approximately 240,000 e-/s/pix with the detector held at 40.5 K in an 18 K environment (Section \ref{sec:det_measured_perf}). Frame rate and operating temperature can shift this value but its impact on the noise will always be small compared to the telescope contribution. The 1/f noise is calculated as described in Section \ref{sec:programtheory}. If a user does not include a chopping and/or nodding frequency, the program uses a separate empirical estimate for the staring mode noise floor of approximately 0.4 ADU/pix. This value does not scale with number of frames. Finally, the average bias is added to the pixels to determine the well-depth of the observation. The well-depth, SNR for a single frame, and required exposure time to achieve a desired overall SNR are all returned by the program. Comparisons of the Alpha Boo data with the program's results after empirical adjustments find the results agree to within 10\%.

    An additional feature of the program assesses the resultant SNR of an attempted measured flux ratio between two filters. The process is identical to the above but during the final SNR calculation, the user instead provides the planned observing time in each filter. The program will estimate the SNR for each filter and then output the expected flux ratio of the two (with an associated uncertainty). This can be used to determine if a significant flux ratio can be measured between different filters such as the narrow-band ammonia feature filter and the wider N' filter.

\section{Discussion}
    \label{sec:discussion}
    \subsection{Performance Limits and Comparisons}
    \label{sec:perfomance_limits}
    Paired with the commissioned MAPS AO, MIRAC-5 will be capable of diffraction limited performance in all bandpasses. We calculate the limiting magnitude for the L', M', and N' filters assuming an ideal Airy disk PSF, an 8 hour observing window at 100\% efficiency, and a required SNR of 5. For N' at a frame rate of 50 Hz, a chopper frequency of 1 Hz is employed to reduce the impact of 1/f noise below telescope plus instrument shot noise. However, for L' at a frame rate of 1 Hz, it is sufficient to nod on an approximately 2 minute cycle and disable the chopper entirely. With the new dichroic, we anticipate we will be able to relax the modulation frequency for M' and that said filter will only require nodding as well. Real observing efficiency for all bands will be on the order of 90\% efficiency. In Table \ref{tab:performance_limits}, we catalog these results for the three bandpasses, considering the current capabilities and an ``upgrade dichroic'' adjustment, i.e., how much would the performance improve once we swap the new dichroic. It is expected with optimized parameters for chop-nod and frame rate that the limiting magnitude could be improved by 0.1 mags in each band (which is not included in Table \ref{tab:performance_limits}). We find that the dominant sources of noise are the telescope plus instrument emission (N' and M') and the ambient dark current (M' and L').

    \begin{table}[]
        \centering
        \begin{tabular}{c|c|c|c|c|c|c}
            Band & Empirical & Frame & Well & Nod or Chop & $e_{obs}$ & Estimated \\
             & Background Adj. &  Rate &  Depth & Modulation && Limiting \\ 
             &  & (Hz) & (\%) & (Hz) & (\%) & Magnitude \\ 
             \hline
             L' & x0.94 & 1 & 24.4 & 1/120 & 97 & 17.4 \\%.122" FWHM
             L' **& x0.94 & 1 & 16.0 & 1/120 & 97 & 18.0 \\%.122" FWHM
             M' & x2.65 & 5 & 37.6 & 1/10 & 93.4 & 14.1 \\%.148" FWHM
             M' **& x2.65 & 5 & 18.4 & 1/10 & 93.4 & 15.6 \\%.148" FWHM
             N' & x0.48 & 50 & 70.1 & 1 & 91.5 & 11.3\\ %.376" FWHM
             N' **& x0.48 & 50 & 30.3 & 1 & 91.5 & 12.6\\
        \end{tabular}
        \caption{A catalog for performance limits for MIRAC-5, assuming a minimum required SNR = 5, science exposure time of 8 hours, and an ideal PSF from the source sampled within the FWHM. A chopper or nodding frequency and resulting efficiency are provided for reference and utilized for 1/f noise contributions. We demonstrate the benefit of the future dichroic in results marked ** to emphasize how performance will improve once the current 50:50 dichroic is replaced. The improvements to throughput and the reduction in noise substantially improve the expected performance.}
        \label{tab:performance_limits}
    \end{table}

    We can compare the MIRAC-5 background limited performance with other ground-based instruments. We focus on background limited performance, leaving contrast performance for analysis following a future coronagraph upgrade. Working from published data, we assume all systems' SNRs scale with 1/sqrt(N) scaling and put them into a common reference frame of 8-hours and SNR = 5. We list these systems below:
    \begin{itemize}
        \item VISIR on VLT with AO (via data from the NEAR project, \cite{wagner_2021}): We begin with Figure 4 from their paper which provides background limited performance at SNR = 3, with the AO enabled, coronagraphic throughput, and an exposure time of 80 hours in N-band of about 0.4 mJy. We convert to a 8-hour reference frame by multiplying by $\sqrt{80/8}$, finding we would require a 1.26 mJy target. Assuming the system is limited by background noise, we can then multiply by $(5/3)^2$ to find the SNR = 5 detection limit in 8 hours: 3.5 mJy or about 9.9 magnitude in N-band. 
        \item VISIR on VLT without AO (via data from the VISIR user manual): We use the background sensitivity for the small field in median observing conditions for the SiC (11.85 micron) filter (as seen in the manual's Table 2). This non-AO supported (but higher throughput) data returns a sensitivity limit of 7 mJy per hour at 10$\sigma$. We convert to our 5$\sigma$ and 8 hour reference frame by multiplying the sensitivity by 0.25 and dividing by $\sqrt(8)$, finding the SNR = 5 detection limit in 8 hours: 0.62 mJy or about 11.7 magnitude in N-band.
        \item NIRC2 on Keck with AO (via data from \cite{bowensrubin2023AJ....166..260B}): The survey achieved a M-band limiting magnitude of 14.38 with 2 hours of exposure time across one night. If a full 8 hours was utilized, the expected M-band limiting magnitude would be 15.1.
        % \item NIRC2 on Keck without AO (via data from the NIRC2 user manual): We utilize the measured sensitivities data for L' and Ms with the narrow camera (0.01"/pix plate scale) and circumscribed pupil for a point source detection (assuming perfect Strehl). The data is given for 5$\sigma$, 1 hour exposures so we convert to the 8 hour reference frame, finding a limiting magnitude in L' of 21.1 and in Ms of 18.6.
        \item NACO on VLT with AO (via analysis of 2004 and 2011 data by \cite{sauter2024PASP..136i5001S}): The analysis scales two sets of NACO on VLT data taken several years apart with slightly different observing parameters (dithering strategy, usage of nodding). The 50 minute detection limits return 5$\sigma$ limits of 17.1 and 17.5 in L band. Using an average of 17.3, we scale to 8 hours, finding a limiting magnitude of 18.5 in L-band. The paper notes that deviations from ideal scaling are already visible at 50 minutes and thus the final background limiting magnitude may be closer to 17.8. The paper determines the impact of AO deformable mirror variations limits the achievable background limit.
        % \item ERIS on VLT \citep{davies2023A&A...674A.207D}: 
        % We utilize ESO's Exposure Time Calculator 2.0 for ERIS, selecting the Lp filter with 0.4 second exposure time in fast readout mode. We find that the expected SNR for a L-mag = 19.2 target is 0.0187 per frame such that we should achieve SNR = 5 with 8 hours of exposure time%using the davies paper, I originally estimated 15.6 for the br-a-cont in 8 hours SNR = 5. I think it's worth noting if we use the calculator, it reports we achieve a limiting mag of about 18. So there is a huge discrepancy with the calculator... It's possible I screwed up the calculation but it is very probable that at least to some extent, the calculator is overly optimistic. That is of course a big point of our paper...
    \end{itemize}
    %Theoretical GeoSnap on LBTI: Also about 13 mag in N-band in 8 hours
    %MIRI on JWST: About 16.9 mag in 4183 seconds? (M-band)

    MIRAC-5's future L', M', and N' performance is similar or favorable compared with other ground-based infrared instruments. Reducing the ambient dark current and telescope plus instrument emission could further improve the MIRAC-5 limiting magnitudes. However, based on discussion in \cite{bowensrubin2023AJ....166..260B}, it is possible several hour integrations may not continuously scale as 1/$\sqrt{t}$ owing to noise sources with non-Poisson distributions. This may be due to shifts in the precipitable water vapor across a night \citep{bowensrubin2023AJ....166..260B} or could be due to systematic fluctuations in the background induced by AO systems \citep{sauter2024PASP..136i5001S}. Another possibility is excessive temporal co-adding of data, i.e., sampling the sky/telescope backgrounds at rates slower than 1 Hz, which may be the cause of departure from 1/$\sqrt{t}$ scaling owing to significant variations in the background at those timescales in the infrared \citep{allen1981PASP...93..381A, absil10.1117/12.549311}. Our synthetic co-adding comparisons in Figure \ref{fig:noise_pair_long} may support this finding and indicate that by avoiding co-adding prior to analysis, the sensitivity for mid-IR instruments can be greatly improved. This is a topic we plan to investigate in future work. MIRAC-5's limiting magnitudes will require reassessment following future improvements and full night integration tests.
    
    \subsection{Current and Future Use Cases}
    %Then "future work" paragraph
    %Coronagraph should be mentioned back in modes section
    MIRAC-5 is already capable of certain science objectives including observations of T Tauri stars or the characterization of Starlink satellites on mid-IR observations. The rapidly increasing number of these satellites may impact future programs such as METIS on ELT. We predict via simple models that these satellites can produce saturated streaks in N-band and we will use MIRAC-5 to explore this. This will then help us estimate the performance impact future satellite constellations will have on ground-based mid-IR systems.
    
    MIRAC-5 is expected to see several improvements. In the short-term, we will swap the current 50:50 dichroic with a new dichroic of significantly higher transmission. At shorter bandpasses such as L-band, we are hampered by the high instrumental background and as such we are working to identify and reduce the contributors to this background. Within the next year, the full commissioning of MAPS AO will allow MIRAC-5 to reach the ideal image qualities estimated in Section \ref{sec:perfomance_limits}. This will immediately open up several key science objectives including observations of wide orbit companions. Unfortunately, originally planned work such as observations of ammonia in GJ504b \citep{bowens2022SPIE12184E..1UB} may not be possible owing to significantly higher background emission than estimated at the time, even following the dichroic upgrade. 

    In the next two years, MIRAC-5 will be upgraded with an AGPM. Supported by a Quadrant analysis of Coronagraphic Images for Tip-tilt Sensing (QACITS) \citep{huby2017A&A...600A..46H}, it should be possible for MIRAC-5 to observe targets within the contrast-limited regime with capabilities comparable with JWST. Potential targets include Eps Indi A b or 51 Eri b, forming protoplanets, and newly discovered long-period RV planets.

\section{Conclusion} \label{sec:conclusion}
    %Michael: 2 to 3 sentence paper summary
    MIRAC-5 is a ground-based mid-infrared imager operating on the 6.5 m MMT telescope. It is well positioned to push ground-based mid-IR observing forward in advance of the ELT era thanks to its state-of-the-art GeoSnap detector and MAPS AO support. We focused on quantification of key instrument parameters before using these to create a self-consistent SNR and exposure time calculator. In short, we found that MIRAC-5 exhibits:
    %Then bulleted list of major quantitative conclusions
    \begin{itemize}
        \item Instrumental dark signal of approximately 2.4E5 e-/s/pix
        \item 1/f noise that matches prior empirical assessments and can be mitigated by spatially modulating images via the built-in chopper and/or telescope nodding
        \item N-band noise dominated by telescope plus instrument emission, with a current RMS value of 6400 e-/pix/s
        \item QE pixel-to-pixel variations on the order of 4\%
        \item Current throughputs including QE between 5\% to 12\%, depending on the specific bandpass (with future throughputs expected to be between 7\% and 40\%)
        \item Current image quality without AO near the diffraction limit, particularly for N'
        \item Expected background limiting magnitudes of 18.0 in L', 15.6 in M', and 12.6 in N' for an 8-hour exposure making it comparable or superior to past ground-based mid-IR systems
    \end{itemize}

    In the short-term, MIRAC-5 will benefit from an upgraded dichroic and the complete commissioning of the MAPS AO system. This will enable testing of long integration performance and open up several science objectives. A future coronagraph upgrade will further enhance MIRAC-5, enabling high contrast imaging that should be comparable to JWST. Mid-IR high contrast imaging done with MIRAC-5 will also inform program choices for future mid-IR ELT imagers such as METIS on ELT \citep{brandl2021Msngr.182...22B}, MICHI on TMT \citep{packham2012SPIE.8446E..7GP}, or TIGER on GMT \citep{hinz2012SPIE.8446E..1PH}. .

    % Improvements to the AO will allow us to better define the image quality
    %     This will enable X near term science
    % MAPS AO's ongoing commissioning will only improve the image quality and other capabilities.

    % Even before a coronagraph, MIRAC-5 can image wide orbit gas giants (GJ504b, citations), protoplanetary disks and objects within, T Tauri...
    % % Coronagraph Upgrade
    % %     Image of pupil?
    % %     This will enable Y long term science

% \section{Software and third party data repository citations} \label{sec:cite}

% PASP would like to encourage authors to change software and
% third party data repository references from the current standard of a
% footnote to a first class citation in the bibliography.  As a bibliographic
% citation these important references will be more easily captured and credit
% will be given to the appropriate people.

% The first step to making this happen is to have the data or software in
% a long term repository that has made these items available via a persistent
% identifier like a Digital Object Identifier (DOI).  Guidance on how to properly cite the software you use in your PASP article can be found in the PASP author instructions at \break \url{https://iopscience.iop.org/journal/1538-3873/page/instructions_for_authors#citation}. More extensive guidance, including a list of repositories
% that satisfy the persistent DOI criteria, plus each one's pros and cons, can be found at \break
% \url{https://github.com/AASJournals/Tutorials/tree/master/Repositories}.

\begin{acknowledgments}
We would like to acknowledge the contributions to this project from Judith Piper$^\dag$ and her work at the University of Rochester. This project was generously funded by the Heising-Simons Foundation. 

We would like to thank the U.S. National Oceanic and Atmospheric Administration (NOAA) for the Legacy Vertical Temperature Profile, Legacy Vertical Moisture Profile and Total Precipitable Water product data from the GOES-16 satellite.
\end{acknowledgments}


\vspace{5mm}
\facilities{MMT}


% \software{astropy \citep{2013A&A...558A..33A,2018AJ....156..123A},  
          % Cloudy \citep{2013RMxAA..49..137F}, 
          % Source Extractor \citep{1996A&AS..117..393B}
          % }



\appendix

\section{Single Frame Mosaics}
\label{ap:singleframe}
\begin{figure}
    \centering
    \includegraphics[width=0.5\linewidth]{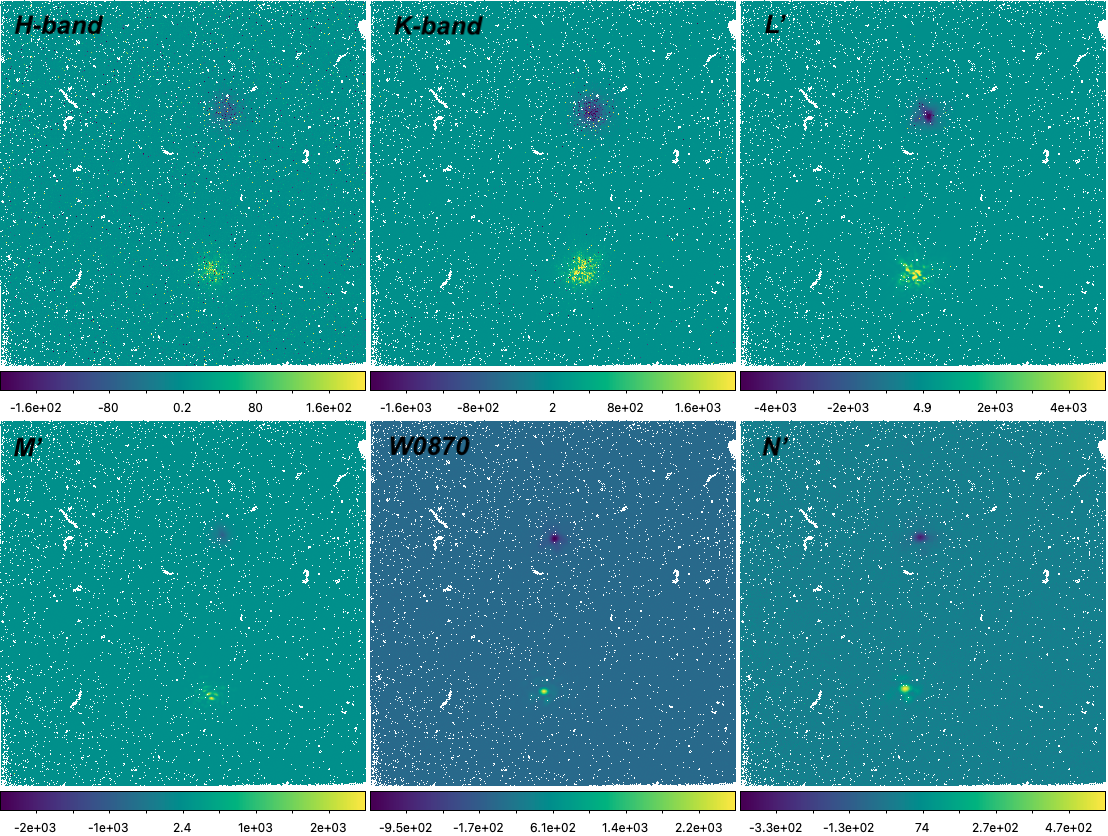}
    \caption{A mosaic of Alpha Boo frames in \textbf{an user-defined scale}} for six key bandpasses (H-band, K-band, L', M', W0870, and N'). The mosaic is constructed from a single nod A frame subtracted by a mean of nod B frames, thus making the positive image of the star the result of a single frame. We have applied a flat field and bad pixel mask as well.
    \label{fig:mos_custom}
\end{figure}

\begin{figure}
    \centering
    \includegraphics[width=0.5\linewidth]{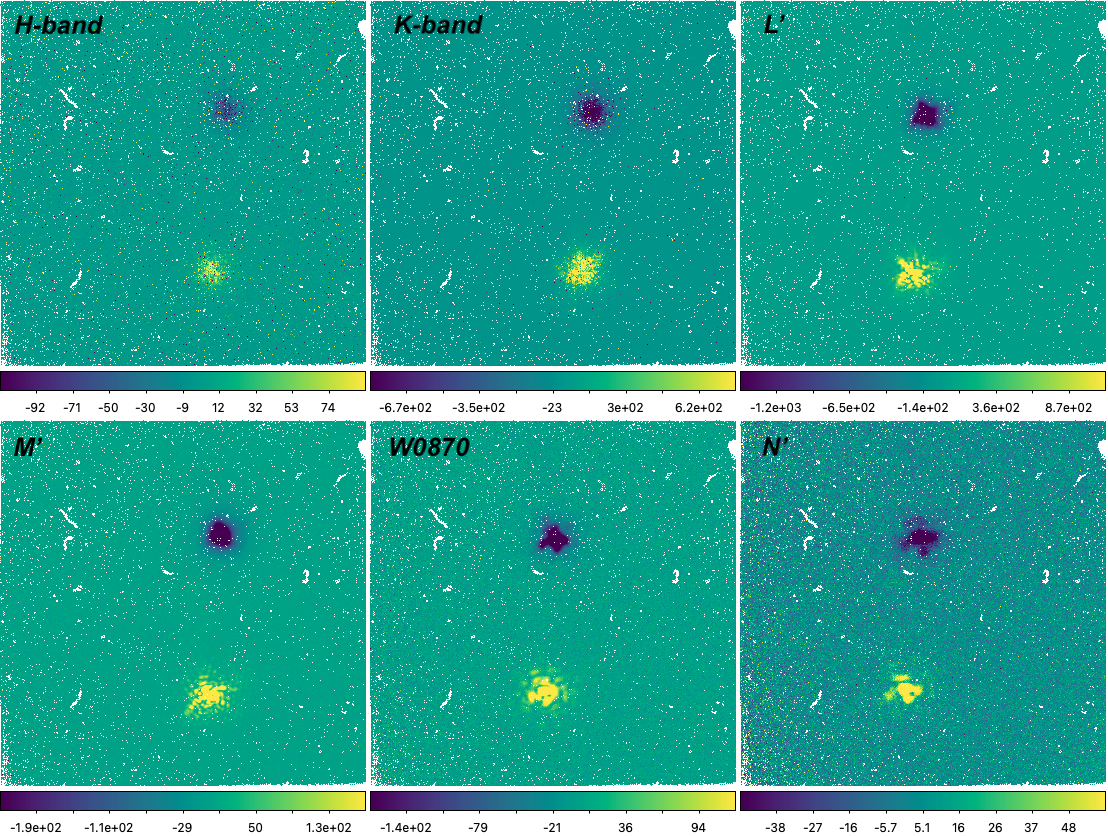}
    \caption{A mosaic of Alpha Boo frames in \textbf{99.5\% scale} for six key bandpasses (H-band, K-band, L', M', W0870, and N'). The mosaic is constructed from a single nod A frame subtracted by a mean of nod B frames, thus making the positive image of the star the result of a single frame. We have applied a flat field and bad pixel mask as well.}
    \label{fig:mos_99p5}
\end{figure}

\begin{figure}
    \centering
    \includegraphics[width=0.5\linewidth]{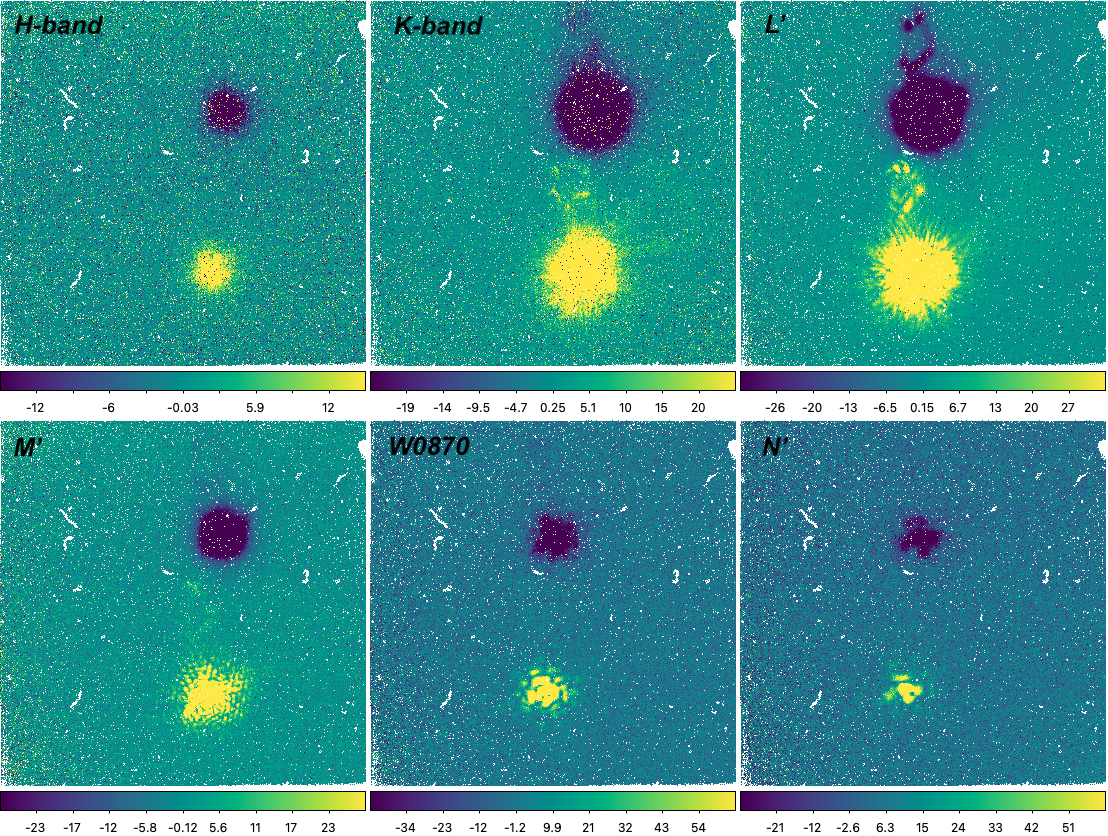}
    \caption{A mosaic of Alpha Boo frames in \textbf{Z-scale} for six key bandpasses (H-band, K-band, L', M', W0870, and N'). The mosaic is constructed from a single nod A frame subtracted by a mean of nod B frames, thus making the positive image of the star the result of a single frame. We have applied a flat field and bad pixel mask as well. A 1\% ghost is visible prominently in the K-band and L' images. We have identified it as originating from somewhere within BLINC.}
    \label{fig:mos_zscale}
\end{figure}

Prior to \textit{PYNPOINT} \citep{amara2012MNRAS.427..948A,stolker2019A&A...621A..59S} reduction, we present the first frames for the nod A position in Figures \ref{fig:mos_custom}, \ref{fig:mos_99p5}, and \ref{fig:mos_zscale}. We have applied some reduction to these frames including subtraction by the mean nod B image, flatfielding, and bad pixel masking in order to make visible Alpha Boo. The positive image of Alpha Boo is representative of a single image quality. The figures are provided with different scales to help highlight features of the image.

\section{Zero Magnitude References}
\label{ap:zeromag}
Utilizing the source flux density found via photometry of Alpha Boo (Section \ref{sec:throughputs}), we calculate zero magnitude flux density for MIRAC-5 in Table \ref{tab:zeromags}. Frame rates and total transmissions from Table \ref{tab:throughputs} are used to convert source intensity into a bandpass flux. We assume uncertainty is dominated by the uncertainty in the source intensity and $\pm$0.03 in magnitude uncertainty \citep{cohen1995AJ....110..275C}.

\begin{table}[!h]
    \centering
    \begin{tabular}{c|c|c|c|c|c|c|c}
    Band     &  Central $\lambda$  & FWHM  & Source & Source Signal & Source Signal& Zero Mag Flux Density  & Uncertainty \\
     & ($\mu$m) & ($\mu$m) & Mag. & (Ge-/s) & SNR & (Jy) & (Jy) \\
    \hline
    L'     & 3.84 & 0.62 & -3.15& 35 & 135 & 180 & 6.3 \\
    M' & 4.66 & 0.24 & -2.93 & 7.2  & 142 & 225 & 11 \\
    N' & 11.34 & 2.27 &-3.14& 4.3 & 110 & 24.3 & 0.89 \\
    \end{tabular}
    \caption{Zero magnitude references derived for MIRAC-5 utilizing Alpha Boo photometry.}
    \label{tab:zeromags}
\end{table}

\newpage

\bibliography{PASPbib}{}
\bibliographystyle{aasjournal}

\end{document}